\documentclass{aa}  

\usepackage{graphicx}

\usepackage{amsmath,amssymb}
\usepackage[flushleft]{threeparttable}
\usepackage{array,booktabs,makecell}
\usepackage{enumitem}
\usepackage{multirow}
\usepackage[dvipsnames]{xcolor}
\usepackage[normalem]{ulem}

\usepackage{color}
\usepackage{xcolor}

\newcommand{\I}{{\cal{I}}}
\newcommand{\SNRfil}{({\rm S/N})_{\rm fil}}

\usepackage{txfonts}
\usepackage[]{hyperref}
\begin{document}

   \title{FilDReaMS \\ 
   1. Presentation of a new method 
   for \textbf{Fil}ament \textbf{D}etection and \textbf{Re}construction \textbf{a}t \textbf{M}ultiple \textbf{S}cales.}
   
   \titlerunning{FilDReaMS. 1. A new method to detect filaments and determine their orientations}

   \author{J.-S. Carrière
          \inst{1}
          \and
          L. Montier\inst{1}
          \and
          K. Ferrière\inst{1}
          \and
          I. Ristorcelli\inst{1}
          }

   \institute{IRAP, Université de Toulouse, CNRS, 9 avenue du Colonel Roche, BP 44346, 31028 Toulouse Cedex 4, France\\
              \email{jeansebastienpaulcarriere@gmail.com}
             }

   %\date{Received September 15, 1996; accepted March 16, 1997}

% \abstract{}{}{}{}{} 
% 5 {} token are mandatory
 
  \abstract
   {Filamentary structures appear to be ubiquitous in the interstellar medium. Being able to detect and characterize them  is the first step toward understanding their origin, their evolution, and their role in the Galactic cycle of matter.}
   {We present a new method, called {\tt FilDReaMS}, to detect and analyze filaments in a given image. This method is meant to be fast, user-friendly, multi-scale, and suited for statistical studies.} 
   {%%Our main tool is a rectangular model bar, which is moved and rotated across the entire image
  The input image is scanned with a rectangular model bar, which makes it possible to uncover structures that can be locally approximated by this bar and to derive their orientations. The bar width can be varied over a broad range of values to probe filaments of different widths.}
    {We performed several series of tests to validate the method and to assess its sensitivity to the level of noise, the filament aspect ratios, and the dynamic range of filament intensities. We found that the method exhibits very good performance at recovering the orientation of the filamentary structures, with an accuracy of $0.5^{\circ}$ in nominal conditions, up to $3^{\circ}$ in the worst case scenario with high level of noise. The filaments width is recovered with uncertainties better than $0.5\,\rm{px}$ (pixels) in most of the cases, which could extend up to $3\,\rm{px}$ in case of low signal-to-noise ratios. Some attempt to build a correspondence between Plummer-type filament profiles and outcomes of the method is proposed, but remains sensitive to the local environment.}
    {Our method is found to be robust and adapted to the identification and the reconstruction  of filamentary structures in various environments, from diffuse to dense medium. It allows us to explore the hierarchical scales of these filamentary structures with a high reliability, especially when dealing with their orientation.}


   \maketitle
%
%-------------------------------------------------------------------

\section{Introduction}
\label{sec:introduction}

A large variety of methods have already been developed to extract elongated structures in two-dimensional (2D) maps. Their approaches may be divided into three main categories: some focus on a purely \textit{local} analysis of the structures, based on local derivatives at each pixel; others adopt a \textit{non-local} approach, by exploring a larger space around each pixel to look for specific scales; the third category of methods propose a \textit{global} analysis of the whole map, applying a multi-scale decomposition. 

\textit{Local} methods compute either the gradient (first-order derivatives) \citep[e.g.,][]{Soler_2013, Soler_2016}
or the Hessian matrix (second-order derivatives) \citep[e.g.,][]{Polychroni_2013, Schisano_2014, Bracco_2016} at each pixel
of the considered map. %%(intensity or column-density) map. 
In some cases, the main purpose is to derive the orientations of elongated structures for statistical purposes \citep[e.g.,][]{Soler_2013, Soler_2016, Bracco_2016}.
In other cases, filament skeletons are extracted by connecting contiguous pixels along the crests of the (intensity or column-density) distribution.
For instance, the {\tt DisPerSe} method, originally developed to recover filament skeletons in cosmic web maps \citep{Sousbie_2011}, was successfully applied to {\it Herschel} column density maps \citep[][]{Arzoumanian_2011, Arzoumanian_2019, Peretto_2012, Palmeirim_2013} and to $^{13}{\rm CO}$ intensity maps \citep{Panopoulou_2014}.
A limitation of the local approach is the difficulty of detecting faint structures such as striations.\footnote{
Here, we use the term striations to refer to the faint and periodic structures seen in the {\it Herschel} maps. These are similar to the periodic magnetic-field-aligned structures detected in the diffuse $^{12}$CO emission from the Taurus molecular cloud \citep{Goldsmith_2008, Narayanan_2008}.}

\textit{Non-local} methods focus on a given scale around each pixel. The template-matching approach \citep{Juvela_2016} allows one to look for any specific morphology of given dimensions and to build the probability of finding such oriented structures through a kernel convolution on the map. Another efficient approach is the Rolling Hough Transform \citep[{\tt RHT},][]{Clark_2014} method, which computes an estimator of the level of linearity of the structures in the neighbourhood of a pixel at a given scale, making use of the Hough transform. This method was extensively used in recent studies, in HI, {\it Herschel}, and {\it Planck} maps \citep[][]{Clark_2015, Clark_2019, Malinen_2016, Panopoulou_2016, Alina_2019}.
A third approach is the {\tt filfinder} method, which extracts filament skeletons \citep{Koch_215}. However, in contrast to {\tt DisPerSe}, {\tt filfinder} starts with a spatial filtering at a given scale and covers a dynamic range broad enough to include striations.

Finally \textit{global} methods offer a multi-scale and complete analysis of a field.
The {\tt getfilaments} method \citep{Men'shchikov_2013} extracts  a filament network with the help of statistical tools and morphological filtering to remove background noise. It also extracts point sources and is able to perform a multi-waveband analysis. This method is very complete, but it requires some fine-tuning and additional tools to extract the filament orientations and scales. It was already applied to {\it Herschel} maps \citep{Cox_2016, Rivera-Ingraham_2016, Rivera_Ingraham_2017, Arzoumanian_2019}.
The wavelet-based methods by \citet{Robitaille_2019} and by \citet{Ossenkopf-Okada_2019} use an anisotropic-wavelet analysis to extract a whole filament network by analyzing fluctuations in the map as functions of spatial scale. 
This kind of method may circumvent the well-known biases of other typically-used methods \citep{Panopoulou_2017} and it remains quite fast. However, it requires additional steps to extract filament orientations. Although this method is multi-scale, the wavelet analysis implies a logspace scaling, which results in a lower resolution at higher scale.

A full comparison between these different methods would be extremely useful, %%but it is still missing today. 
but to date only a few limited comparisons exist.
For instance, \citet[][]{Juvela_2016} compared the template-matching and {\tt RHT} methods applied to simulation data. He found that both methods give equally good results, except in simulations with significant noise and background fluctuations, for which template matching performs better. 
More recently, \citet{Micelotta_2021} compared the gradient and {\tt RHT} methods, again on simulation data.
They found similarities, but also disparities, in the results, and they attributed the disparities to intrinsic differences in the filamentary structures selected by both methods.

Here, we would like to study the relative orientations between filaments and the local magnetic field.
To that end, we need a filament extraction method that can operate in a broad range of Galactic environments, from dense and complex structures to the more diffuse (neutral) medium, and can provide robust and homogeneous filament detections in various fields for a multi-scale statistical analysis. While none of the methods described above is fully satisfactory for our purpose, the closest is probably {\tt RHT}, which is easy to use and has already given promising results \citep[see, e.g.,][]{Clark_2014, Malinen_2016, Panopoulou_2016, Alina_2019}. Therefore, we decided to start from {\tt RHT} and adapt it %KF we decided to adapt {\tt RHT} KFand develop a new method 
to match our requirements.
This led us to develop a new method, called {\tt FilDReaMS} ({\bf Fil}ament {\bf D}etection and {\bf Re}construction {\bf a}t {\bf M}ultiple {\bf S}cales).

In Sect.~\ref{sec:overview}, we present the main features of our new {\tt FilDReaMS} method, with reference to {\tt RHT}.
In Sect.~\ref{sec:Methodology}, we provide a detailed description of the {\tt FilDReaMS} methodology.
In Sect.~\ref{sec:Validation}, we present the results of several series of simulations designed to validate {\tt FilDReaMS}.
In Sect.~\ref{sec:comparison}, we apply {\tt FilDReaMS} to the {\it Herschel} G210 field and compare our results to those previously obtained with {\tt RHT}.
In Sect.~\ref{sec:conclusion}, we conclude our paper.

\section{General overview}
\label{sec:overview}

\subsection{Introduction to \texttt{FilDReaMS}}
\label{sec:FilDReaMS_intro}

Let us consider a map of a given quantity $\I$, which, in the astrophysical context, can represent intensity, column density, or temperature. This map will be our reference to present the {\tt FilDReaMS} method. For simplicity, we will refer to $\I$ as being an intensity, keeping in mind that $\I$ actually has a broader meaning.

The purpose of applying {\tt FilDReaMS} to the map of $\I$ is to identify filamentary structures over a range of spatial scales and intensities, from the largest and brightest filaments down to striations. By considering that filaments can be locally approximated by rectangular bars, {\tt FilDReaMS} is able to extract two of their characteristics: the widths and the orientations of the associated bars.

In the following, the rectangular bar used by {\tt FilDReaMS} is referred to as the model bar.
It is characterized by a width $W_{\rm b}$, a length $L_{\rm b}$, and an aspect ratio $r_{\rm b} = L_{\rm b} / W_{\rm b}$.
For any filament detected with a model bar of width $W_{\rm b}$, $W_{\rm b}$ is referred to as the bar width of the filament.

The orientation angle of the model bar, $\psi_{\rm b}$, is defined with respect to a given north direction (e.g., Galactic north for an astrophysical map) and taken to increase counterclockwise from north.
This definition is consistent with the IAU convention for polarisation angles.
Since the model bar is symmetric, $\psi_{\rm b}$ can be defined  over a $180^{\circ}$ range, which we choose to be $[-90^{\circ}, +90^{\circ}]$.

We adopt the same convention for the orientation angle of a filament, $\psi_{\rm f}$, defined in Sect.~\ref{sec:orientation_angle}.
Furthermore, when considering the difference between two angles defined in $[-90^{\circ}, +90^{\circ}]$, we require the result to also lie in the range $[-90^{\circ}, +90^{\circ}]$, possibly by adding or subtracting $180^{\circ}$.

All the parameters related to {\tt FilDReaMS}
are listed in Table~\ref{tab:FilDReaMS_notations}.

\begin{table*}
\caption{List of all the symbols used in the paper.}
\centering
\begin{threeparttable}
\begin{tabular}{m{3.0cm} m{12.5cm}}
\midrule\midrule
A & Initial map of intensity $\I$\\
\cmidrule(l  r ){1-2}
$\I$ & Intensity in a broad sense (intensity, column density, or temperature)\\
\cmidrule(l  r ){1-2}
B & Smoothed map resulting from the convolution of A with a 2D top-hat kernel of radius $R$\\
\cmidrule(l  r ){1-2}
$R$ & Radius of the 2D top-hat kernel\\
\cmidrule(l  r ){1-2}
C & Binary map derived from A$-$B\\
\cmidrule(l  r ){1-2}
$i$ & Index of the considered pixel of map C for the detection of bar-like filaments\\
\cmidrule(l  r ){1-2}
$W_{\rm b}$ & Width of the model bar\\
\cmidrule(l  r ){1-2}
$L_{\rm b}$ & Length of the model bar\\
\cmidrule(l  r ){1-2}
$r_{\rm b} = L_{\rm b}/W_{\rm b}$ & Aspect ratio of the model bar\\
\cmidrule(l  r ){1-2}
$\psi_{\rm b}$ & Orientation angle of the model bar\\
\cmidrule(l  r ){1-2}
$\psi_{\rm f}$ & Orientation angle of a filament\\
\cmidrule(l  r ){1-2}
$f^{\rm M}$ & Measured orientation function\\
\cmidrule(l  r ){1-2}
$\sigma_f$ & Median absolute deviation of $f^{\rm M}$\\
\cmidrule(l  r ){1-2}
$f^{\rm I}$ & Ideal orientation function\\
\cmidrule(l  r ){1-2}
$\Delta\psi$ & Width of the angular window over which $f^{\rm M}$ is compared to $f^{\rm I}$\\
\cmidrule(l  r ){1-2}
$\chi_{\rm r}^{\rm M}$ & Measure of the normalized difference between $f^{\rm M}$ and $f^{\rm I}$ (Eq.~\ref{eq:delta})\\
\cmidrule(l  r ){1-2}
$\I_0$ & Central intensity of synthetic filaments\\
\cmidrule(l  r ){1-2}
$j$ & Index of the considered pixel of map A in one Monte-Carlo iteration\\
\cmidrule(l  r ){1-2}
$\sigma_{{\rm A}_j}$ & Standard deviation of sub-region A$_j$ of map A, centered on pixel $j$\\
\cmidrule(l  r ){1-2}
$\SNRfil = \I_0 / \sigma_{{\rm A}_j}$ & Signal-to-noise ratio of the ideal filament in the Monte-Carlo simulations\\
\cmidrule(l  r ){1-2}
$f^{\rm S}$ & Synthetic orientation function\\
\cmidrule(l  r ){1-2}
$\chi_{\rm r}^{\rm S}$ & Measure of the normalized difference between $f^{\rm S}$ and $f^{\rm I}$\\
\cmidrule(l  r ){1-2}
$(\chi_{\rm r})_{\rm th}$ & Statistical threshold on $\chi_{\rm r}$\\
\cmidrule(l  r ){1-2}
$S = (\chi_{\rm r})_{\rm th}$/$\chi_{\rm r}$ & Significance of filament detection\\
\cmidrule(l  r ){1-2}
C' & Binary map composed of the model bars associated with all the significant filaments\\
\cmidrule(l  r ){1-2}
R & Map of reconstructed filaments\\
\cmidrule(l  r ){1-2}
$i'$ & Index of the considered pixel of map R\\
\cmidrule(l  r ){1-2}
$\psi_{\rm f}^{\star}$ & Orientation angle of the most significant filament\\
\cmidrule(l  r ){1-2}
$W_{\rm b}^{\star}$ & bar width of the most significant filament\\
\cmidrule(l  r ){1-2}
$\sigma_{\mathcal{W}}$ & Standard deviation of white noise in the sets of simulations\\
\cmidrule(l  r ){1-2}
$\sigma_{\mathcal{B}}$ & Standard deviation of Brownian noise in the sets of simulations\\
\cmidrule(l  r ){1-2}
$N_{\rm pix}$ & Number of pixels whose most significant bar width is $W_{\rm b}^\star$\\
\cmidrule(l  r ){1-2}
$N_{\rm map}$ & Total number of pixels in the map\\
\cmidrule(l  r ){1-2}
$W_{\rm b}^{\star{\rm peak}}$ & Most prevalent bar width for the entire map\\
\midrule\midrule
\end{tabular}
\end{threeparttable}
\label{tab:FilDReaMS_notations}
\end{table*}

To provide a first application of {\tt FilDReaMS}, we selected one of the {\it Herschel} fields observed in the Galactic cold core (GCC) key-program \citep[]{Juvela_GCCI_2010, Juvela_GCCIII_2012}, the so-called G210 field, which corresponds to the high Galactic latitude star-forming region L1642. This region was in particular studied in detail by \citet{Malinen_2016}, who investigated the relative orientations between the magnetic field (traced with {\it Planck} polarisation data) and filaments extracted from the G210 {\it Herschel} map using the {\tt RHT} method.
G210 is located at Galactic longitude $l = 210.90^{\circ}$, Galactic latitude $b = -36.55^{\circ}$, and distance $d = 140\pm20\,{\rm pc}$ \citep{Montillaud_2015}.
Its ${\rm H_2}$ column density map has dimensions $\Delta l \times \Delta b = 1.28^{\circ}\times1.22^{\circ}$, corresponding to $3.1\,{\rm pc}\times3.0\,{\rm pc}$.
The angular size of a pixel, equal to one-third of the beam size, is $12"$, corresponding to $0.0081\,{\rm pc}$.
The ${\rm H_2}$ column density, $N_{\rm H_2}$, varies over the range $[0.2, 7.5]\times10^{21}\,{\rm cm}^{-2}$.

\subsection{Overview of \texttt{RHT}}
\label{sec:RHT_overview}

The {\tt RHT} method developed by \citet{Clark_2014} is based on the Hough transform designed to search for straight lines in images, even when partially filled or dominated by noise. Defining a straight line with two parameters (the orientation $\theta_{\tt H}$ of its normal and the minimal distance $\rho_{\tt H}$ to the origin), the Hough transform allows us to pass from the pixel space ($x,y$) to this other parameter space ($\rho_{\tt H}$, $\theta_{\tt H}$), by quantifying the number of pixels located on the same linear feature. In the case of the {\tt RHT} method, in order to focus on straight features and to suppress large-scale structures, a binary version of the original image is first built by subtracting the smoothed image with a top-hat and thresholding to zero. The Hough transform is then applied locally on this binary image: for each pixel, a simplified version of the Hough transform can be applied (with $\rho_{\tt H}$=0) inside a circular area defined around this central pixel in order to pass from pixel space to ($\theta_{\tt H}$) space, and to build the distribution of orientations of the linear features around this central pixel. Once applied iteratively on every pixel of the whole image, the last step of the method consists in choosing a common threshold over which the distributions of orientations are stored, and used to derive local average orientation or total {\tt RHT} intensity over the whole image. 

This method is extremely powerful to get an estimate of the linearity level inside local regions of images irrespective of the overall brightness of the region. It still suffers from a few limitations. Firstly, the use of the Hough transform, which performs transformation from pixel space to ($\rho_{\tt H}$, $\theta_{\tt H}$) space, directly implies that {\tt RHT} distributions of orientation are dependent of the pixelisation of the image, which could lead to subtle bias of the results. Secondly, the choice of the threshold used to cut the distributions of orientations is arbitrary and may impact the analyses from one image to another.

\subsection{\texttt{RHT} versus \texttt{FilDReaMS}}
\label{sec:RHT_FilDReaMS}

{\tt FilDReaMS} tries to overcome some of the limitations of {\tt RHT}, as explained in the rest of this section. It also makes it possible to access additional information about the widths (more exactly, the bar widths) of the detected filaments.

The sensitivity of the algorithm to pixelisation is inherent in the basic implementation of the RHT, which uses relative positions of pixel centers to a given pixel (centered on $x,y$) to build a $\theta_{\tt H}$ representation. {\tt FilDReaMS} solves this problem by a fundamental change of philosophy: it starts from a discretization of the $\theta_{\tt H}$ space to compute the intersection of a rotated bar with any pixel area centered on ($x,y$).

In {\tt FilDReaMS}, the arbitrary choice of threshold in {\tt RHT} for determining linearity significance is alleviated by a comparison to a random distribution obtained locally using ideal template bars. This process takes into account the noise level and the complexity of the region, and provides us a robust assessment of the reliability of the detections. 

The determination of a preferred orientation based on the orientation distribution in each pixel is improved in {\tt FilDReaMS} by performing a search for local maxima, allowing us to derive multiple peaks in the orientation distribution with individual significances, instead of averaging the orientation angle estimated over the whole orientation distribution.

\subsection{The main steps of \texttt{FilDReaMS}}
\label{sec:overview_RHT}

In brief, the main steps of {\tt FilDReaMS} are the following:
\begin{enumerate}
\item {\bf Spatial filtering} : a 2D top-hat filtering is applied to the input image, which is then converted to a binary map that contains only structures narrower than a given width,
see Sect.~\ref{sec:binary_map}.\\

\item {\bf Building an orientation distribution}: the matching between the binary map and 
%%a set of model bars with given width $W_{\rm b}$ and various orientations 
a model bar with given width $W_{\rm b}$ and variable orientation
is evaluated at each pixel $i$, in order to build an orientation distribution, see Sect.~\ref{sec:histogram_of_orientation}.\\

\item {\bf Detection of preferred orientations}: local maxima %%are spotted 
in the orientation distribution at each pixel $i$ are identified with preferred orientations, see Sect.~\ref{sec:orientation_angle}.\\

\item {\bf Reliability assessment}: the significance of each preferred orientation is assessed by comparison with an ideal orientation distribution, see Sect.~\ref{sec:significance}.\\

\item {\bf Reconstruction of physical filaments}: the true shape and the intensity of physical filaments is reconstructed from the initial image masked by the binary map, and a filament orientation angle at each pixel $i'$, $(\psi_{\rm f}^{\star})_{i'}$, is derived, see Sect.~\ref{sec:filament_visualisation}.\\

\item {\bf Iteration over various bar widths}: the procedure is repeated for a range of values of $W_{\rm b}$ in order to derive the most significant bar width at each pixel $i'$, $(W_{\rm b}^\star)_{i'}$, as well as the most prevalent bar widths for the entire map, $W_{\rm b}^{\star{\rm peak}}$, see Sect.~\ref{sec:signif_bar_width}.
\end{enumerate}

While step 1 above is fully similar to the {\tt RHT} processing, steps 2 and 3 have the same objectives as {\tt RHT} but a totally different implementation, allowing to address the pixelisation sensitivity (see Sect.~\ref{sec:histogram_of_orientation}) and the multiplicitiy of the local preferred orientations in case of crossing of linear structures (see Sect.~\ref{sec:orientation_angle}). Finally steps 4 to 6 are entirely new and specific to {\tt FilDReaMS}.

\begin{figure*}
    \centering
    \includegraphics[width=\textwidth]{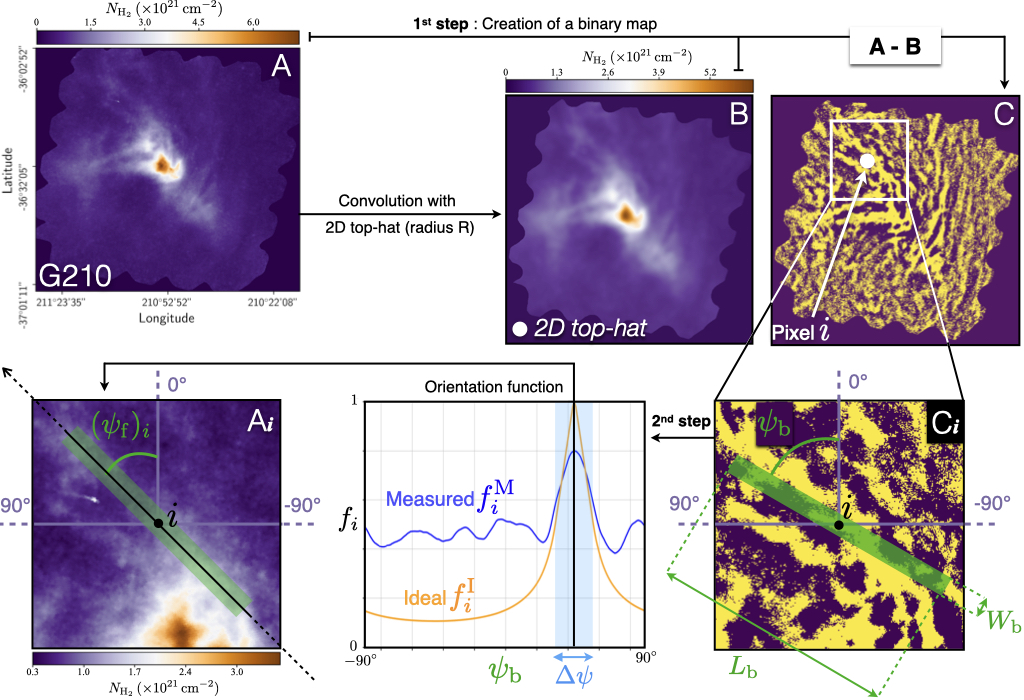}
    \caption{Illustration of the {\tt FilDReaMS} method applied to one pixel $i$ of the ${\rm H_2}$ column density map of the {\it Herschel} G210 field. {\bf Top Left:} Initial map A of the entire G210 field. {\bf Top Middle:} Smoothed map B obtained by convolving map A with a 2D top-hat kernel. {\bf Top Right:} Corresponding binary map C in which the yellow pixels have a value of 1 and the dark pixels a value of 0. {\bf Bottom Right:} Sub-region C$_i$ of map C centered on pixel $i$, with the model bar (width $W_{\rm b}$, length $L_{\rm b}$, and orientation angle $\psi_{\rm b}$) overplotted in green. {\bf Bottom Middle:} Measured orientation function, equal to the fraction of the bar area covered by yellow pixels, $f^{\rm M}_i$, as a function of $\psi_{\rm b}$ (blue curve).
    Also shown are the ideal orientation function, $f^{\rm I}_i$ (orange curve; see Sect.~\ref{sec:significance}) and the angular window of width $\Delta\psi$ (Eq.~\ref{eq:comparison_window}). {\bf Bottom Left:} Sub-region A$_i$ of map A centered on pixel $i$, with the model bar overplotted in green at the orientation angle $(\psi_{\rm f})_i$ of a potential filament (corresponding to the peak of $f^{\rm M}_i$).}
    \label{fig:FilDReaMS_method}
\end{figure*}

\section{Detailed methodology}
\label{sec:Methodology}

In this section, we describe in more detail the successive steps of the {\tt FilDReaMS} method applied to a map of intensity $\I$. As explained at the beginning of Sect.~\ref{sec:FilDReaMS_intro}, the word "intensity", used in connection with the symbol $\I$, is to be understood in a broad sense, which includes quantities such as column density and temperature. To illustrate the method, we provide detailed figures that rely on an ${\rm H_2}$ column density ($N_{\rm H_2}$) map of the {\it Herschel} G210 field.

\subsection{Spatial filtering}
\label{sec:binary_map}

Let us start with a map of intensity $\I$, which we will refer to as the initial map A (top left panel of Fig.~\ref{fig:FilDReaMS_method}).
Our purpose is to identify in this map filaments that can be locally approximated by a rectangular bar of width $W_{\rm b}$.

The first step of {\tt FilDReaMS} is to filter out structures wider than $W_{\rm b}$.
The spatial filtering is performed with the help of a 2D top-hat kernel of radius $R$, whose value is adjusted to the value of $W_{\rm b}$ in the manner explained after the next paragraph.

To begin with, the initial map A is convolved with the 2D top-hat kernel to produce a smoothed map B (top middle panel of Fig.~\ref{fig:FilDReaMS_method}).
Roughly speaking, this smoothing removes structures with widths smaller than $\sim 2R$.
The smoothed map B is then subtracted from the initial map A to produce a map A$-$B from which structures with widths larger than $\sim 2R$ are removed.
Finally, the map A$-$B is transformed into a binary map C by setting all the pixels with positive values to 1 (yellow pixels in the right panels of Fig.~\ref{fig:FilDReaMS_method}) and all the pixels with negative values to 0 (dark pixels).

The adjustement of $R$ to $W_{\rm b}$ is done iteratively by considering increasing values of $R$, starting at $R = 2\,{\rm px}$. For each value of $R$, the initial map A is convolved with a kernel of radius $R$ (as explained above), and the binary map C is examined in search of regions of non-zero pixels wider than $W_{\rm b}$. In practice, this is done by convolving C with a 2D top-hat kernel of diameter $(W_{\rm b}+1\,{\rm px})$; when the normalized convolution reaches a value of 1 at any pixel, we may conclude that this pixel is the center of a disk of diameter $(W_{\rm b}+1\,{\rm px}$) filled with non-zero pixels, which in turn implies that C contains a region of non-zero pixels wider than $W_{\rm b}$. Hence the iteration process stops.

Thus, the binary map C represents the contrasted structures (yellow in Fig.~\ref{fig:FilDReaMS_method}) from the initial map A that are not wider than $W_{\rm b}$.

The convolution gives rise to border effects, with a blank band of width $R$ adjacent to the border of the convolved map B. As a result, map B is somewhat smaller than the initial map A. For large values of $W_{\rm b}$, the blank band may represent a significant fraction of map A; in that case, large filaments close to the edge of map A may escape detection.

\subsection{Orientation distribution}
\label{sec:histogram_of_orientation} 

Let us consider a given pixel $i$ in the full binary map C (top right panel of Fig.~\ref{fig:FilDReaMS_method}) as well as the surrounding sub-region C$_i$ of size $L_{\rm b}$, centered on pixel $i$ (bottom right panel).
Obviously, pixel $i$ must be more distant than $L_{\rm b}/2$ from the border of map C.
Our purpose is to find structures in C$_i$ that can be locally matched to a model bar of width $W_{\rm b}$. 

We consider values of the orientation angle of the model bar, $\psi_{\rm b}$ (defined with the conventions described in Sect.~\ref{sec:FilDReaMS_intro}), spanning the range -90$^{\circ}$ to +90$^{\circ}$ in 1$^{\circ}$ steps. For each value of $\psi_{\rm b}$, the model bar is centered on the considered pixel $i$, and we measure the fraction $f_i\,(\psi_{\rm b})$ of 
the bar area covered by non-zero pixels
(yellow pixels). This gives us the "measured" orientation function, $f^{\mathrm{M}}_i$ (blue curve in the bottom-middle panel of Fig.~\ref{fig:FilDReaMS_method}).

The computation of the area resulting of the intersection of every pixel and the model bar rotated by $\psi_{\rm b}$ is performed only once at the beginning of the processing for all pixels of the $L_{\rm b} \times L_{\rm b}$ domain, and obtained through a numerical drizzling approach. In practice each original pixel is sub-divided into $N_{\rm{drizz}}$ sub-pixels, allowing us to compute an estimate of the exact intersection of the bar and the pixel at a desired accuracy. We adopt a value of $N_{\rm{drizz}}$=101, scaling with $5/W_{\rm b}$, so that we keep an accuracy better than 0.2\% on the computation of the intersecting area with the model bar.  
Once computed, this kernel defined over $L_{\rm b} \times L_{\rm b}$ can be translated at any pixel $i$ location and multiplied with the full binary map to obtain the "measured" orientation function for this pixel.

\subsection{Detection of potential bar-like filaments}
\label{sec:orientation_angle}

\begin{figure}
    \centering
    \includegraphics[width=\columnwidth]{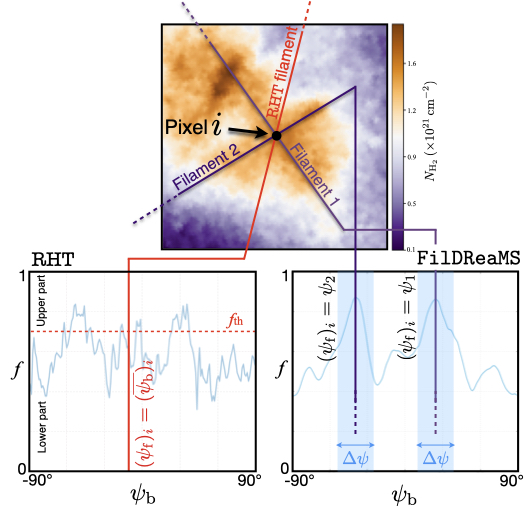}
    \caption{Illustration of a case where the considered pixel $i$ lies at the intersection of two filaments with different orientations. {\bf Top:} Sub-region of the initial map A, highlighting the filaments detected with {\tt FilDReaMS} (dark blue lines) and when computing expectation over {\tt RHT} orientation function (red line). {\bf Bottom:} Orientation function obtained with {\tt RHT} ({\bf Left}) and {\tt FilDReaMS} ({\bf Right}), showing the filament orientation angles derived in both cases.}
    \label{fig:RHT_vs_FilDReaMS}
\end{figure}

The measured orientation function, $f^{\rm M}_i$, makes it possible to detect potential filaments of bar width $W_{\rm b}$ centered on pixel $i$ and to estimate their orientation angle, $(\psi_{\rm f})_i$. We stress that the measured orientation function is relatively smooth over the range of orientation angle, since it results from a kind of convolution with an extended bar at each orientation angle value, so that the determination of local maxima is not affected by local pixel-scale fluctuations (see bottom-right panel of Fig.~\ref{fig:RHT_vs_FilDReaMS}).

{\tt FilDReaMS} identifies the local maxima of the measured orientation function, $f^{\rm M}_i$, through an iterative process. It assigns the first peak to the maximum value of $f^{\rm M}_i$ over the whole range of orientation angle. It then assigns for this peak, an angular window of width $\Delta\psi$, centered on the peak orientation angle, and corresponding to the expected domain of angular extension of the model bar, defined as twice the angle between the bar's diagonals:
\begin{equation}
    \label{eq:comparison_window}
    \Delta\psi = 4 \ \arctan \left( \frac{1}{r_{\rm b}} \right) \, ,
\end{equation}
where $r_{\rm b} = L_{\rm b}/W_{\rm b}$ is the aspect ratio of the model bar.

\noindent
Once the first peak is identified with its own angular window, {\tt FilDReaMS} looks for any other local maximum over the remaining angular domain of the original interval, and assigns to this second peak the corresponding angular orientation and its $\Delta\psi$ angular window. This procedure is repeated until all the local maxima in $f^{\rm M}_i$ are identified. In the event that two consecutive peaks, at $\psi_1$ and $\psi_2$, are so close that their windows overlap ($\vert \psi_1 - \psi_2 \vert < \Delta\psi$), they are considered to actually be part of one and the same potential filament; in that case, only the higher peak is retained together with its window, while the lower peak is ignored. All the peaks remaining after the handling of overlapping windows correspond to potential filaments, whereas the other peaks are considered to be part of the background.

In the case of Fig.~\ref{fig:RHT_vs_FilDReaMS}, {\tt FilDReaMS} detects two potential filaments with orientation angles $\psi_1$ and $\psi_2$ (bottom-right panel), while the common practice with {\tt RHT} orientation functions is to compute the expectation over a given threshold leading to a single preferred orientation in this specific case (bottom-left panel).

\subsection{Reliability assessment of potential bar-like filaments}
\label{sec:significance}

\subsubsection{The significance criterion}
\label{sec:significance_criterion}

\begin{figure*}
    \centering
    \includegraphics[width=\textwidth]{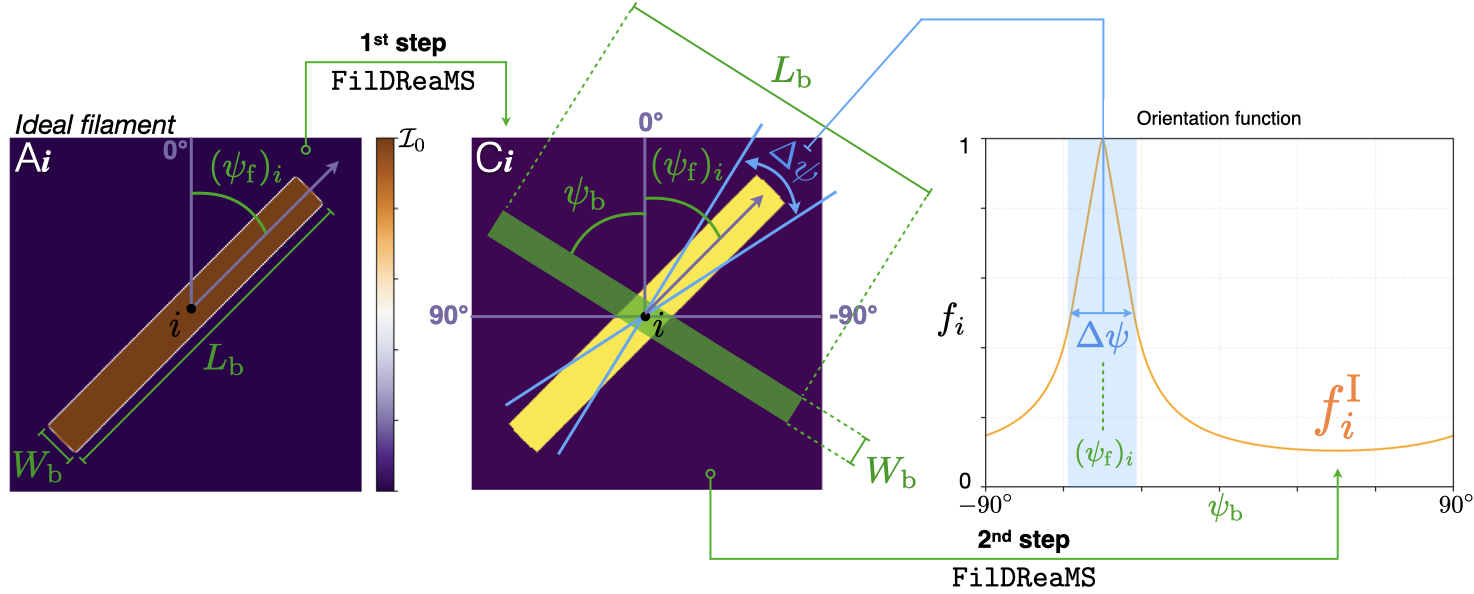}
    \caption{Illustration of one ideal filament and its ideal orientation function.
    {\bf Left:} Map of an ideal filament (filament having the exact same shape as the model bar) centered on pixel $i$ and oriented at angle $(\psi_{\rm f})_i$ (here ($(\psi_{\rm f})_i < 0$).
    {\bf Middle:} Corresponding binary map, with the model bar overplotted in green.
    {\bf Right:} Ideal orientation function, $f^{\rm I}_i$, from the second step of {\tt FilDReaMS} applied to the binary map.}
    \label{fig:Ideal_hist_window}
\end{figure*}

To test the reality of a potential filament detected at the considered pixel $i$, we compare the measured orientation function, $f^{\rm M}_i$, to the ideal orientation function, $f^{\rm I}_i$ (right panel of Fig.~\ref{fig:Ideal_hist_window}), that would be obtained for an ideal filament -- similar to a filament having the exact same shape as the model bar -- superposed on an empty background, at the orientation angle $(\psi_{\rm f})_i$ of the potential filament (left panel). Clearly, at $\psi_{\rm b}=(\psi_{\rm f})_i$, the model bar coincides exactly with the ideal filament, such that all the pixels of the model bar have a value of 1 in the binary map of the ideal filament (middle panel).
As a result, $f^{\rm I}_i$ reaches its peak value, $f^{\rm I}_i = 1$, at $\psi_{\rm b}=(\psi_{\rm f})_i$ (right panel).
Quite expectedly, the width of the peak in $f^{\rm I}_i$ is $\sim \Delta\psi$ (Eq.~\ref{eq:comparison_window}).

{\tt FilDReaMS} then compares the measured and ideal orientation functions over an angular window of width $\Delta\psi$ centered on the peak associated with the potential filament.
This window is both broad enough to capture the relevant characteristics of the potential filament and narrow enough to avoid contamination by a nearby filament at a slightly different angle. The comparison is performed with the help of the parameter

\begin{equation}
\label{eq:delta}
    (\chi_{\rm r}^{\rm M})_i = \sqrt{\frac{\chi^2}{N_\psi}} = \sqrt{\frac{1}{N_\psi} \ \sum^{(\psi_{\rm f})_i + \Delta\psi/2}_{\psi_{\rm b} = (\psi_{\rm f})_i - \Delta\psi/2}\frac{\left( f^{\rm I}_i\,(\psi_{\rm b}) - f^{\rm M}_i\,(\psi_{\rm b}) \right)^2}{\sigma_f^2}}\,,
\end{equation}

\noindent where

\begin{equation}
    \label{eq:MAD}
    \sigma_f = \mathrm{median}\left(\left|f^{\rm M}_i\,(\psi_{\rm b}) - \mathrm{median}\left(f^{\rm M}_i\,(\psi_{\rm b})\right)\right|\right)\,,
\end{equation}

\begin{figure*}
    \centering
    \includegraphics[width=\textwidth]{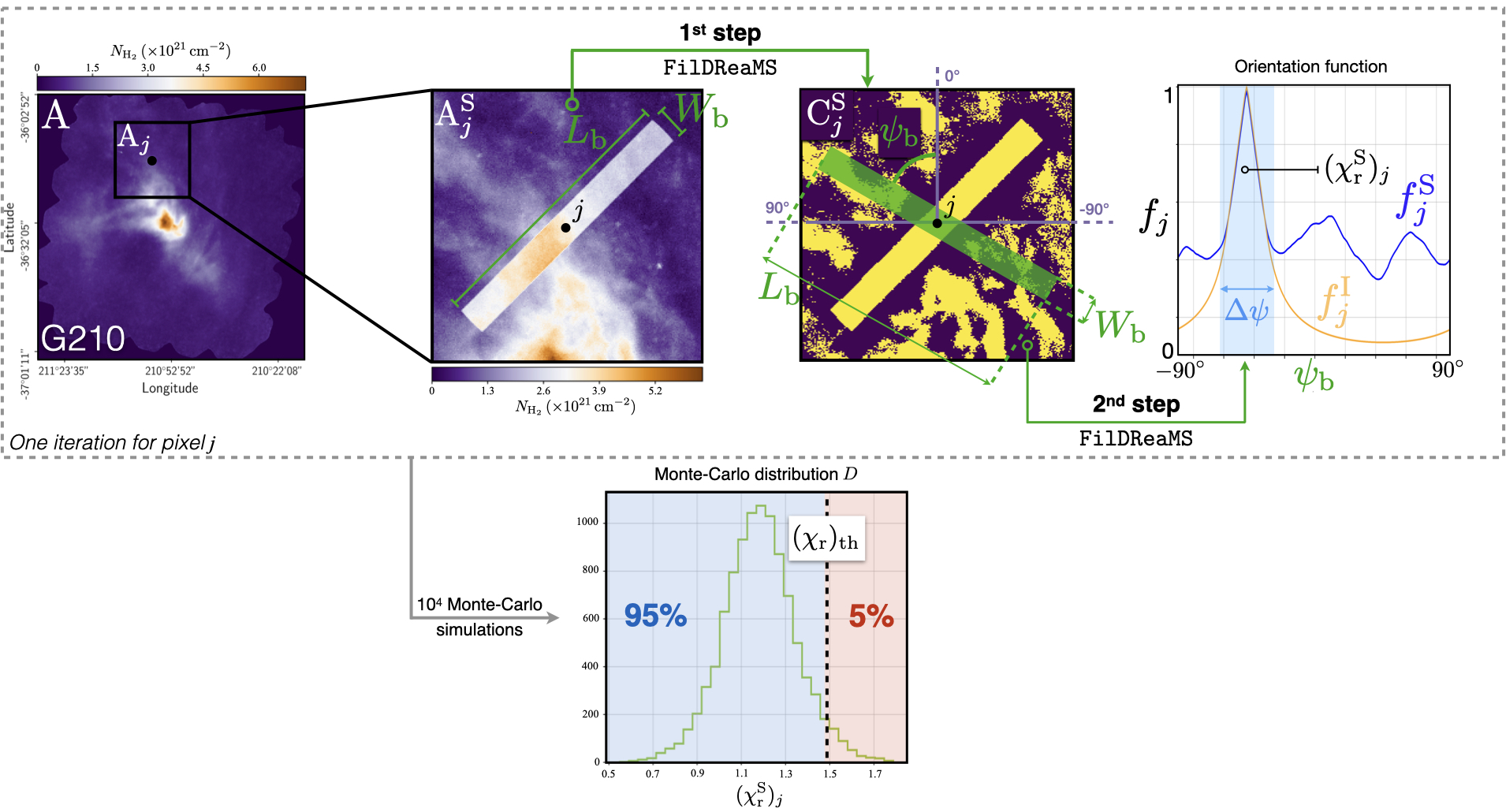}
    \caption{Illustration of %%one Monte-Carlo iteration 
    our Monte-Carlo simulations for the computation of the threshold $(\chi_{\rm r})_{\rm th}$ for a given value of the bar width, $W_{\rm b}$.
    {\bf Leftmost:} Initial map A of the G210 field, with a sub-region A$_j$ centered on an arbitrary pixel $j$.
    {\bf Middle left:} Synthetic map, ${\rm A}^{\rm S}_j$, composed of sub-region A$_j$ plus an ideal filament centered on pixel $j$.
    {\bf Middle right:} Binary map, ${\rm C}^{\rm S}_j$, from the first step of {\tt FilDReaMS} applied to ${\rm A}^{\rm S}_j$, with the model bar overplotted in green.
    {\bf Rightmost:} Synthetic orientation function, $f^{\rm S}_j$ (blue curve), from the second step of {\tt FilDReaMS} applied to ${\rm C}^{\rm S}_j$, together with the ideal orientation function, $f^{\rm I}_j$ (orange curve).
    {\bf Bottom:} Monte-Carlo distribution $D$ of the parameter $(\chi_{\rm r}^{\rm S})_j$ that compares $f^{\rm S}$ and $f^{\rm I}$ within the angular window of width $\Delta\psi$ (see Eq.~\ref{eq:delta} with %%$f^{\rm M}$ replaced by $f^{\rm S}$
    superscript M replaced by S), and threshold $(\chi_{\rm r})_{\rm th}$ corresponding to a $p$-value of 5$\,\%$.
}
    \label{fig:binary_image_significance}
\end{figure*}

\noindent $N_\psi$ is the number of $\psi_{\rm b}$ values within the comparison window $\Delta\psi$, and $\sigma_f$ is the intrinsic error in $f^{\rm M}_i\,(\psi_{\rm b})$ defined as the median absolute deviation computed over all the angles $\psi_{\rm b}$ of all the pixels $i$ of the initial map A. $(\chi_{\rm r}^{\rm M})_i$, which is actually similar to the square root of a reduced chi-squared, quantifies how close a potential filament is to a rectangular bar in an empty background.
 
We assess the significance of the potential filament by comparing $(\chi_{\rm r}^{\rm M})_i$ to a threshold, $(\chi_{\rm r})_{\rm th}$, derived through Monte-Carlo simulations. Since these simulations need to be run only once (for any given $W_{\rm b}$) for the entire map A, not for each pixel $i$ separately, we describe them in a separate subsection (Sect.~\ref{sec:montecarlo}).

We consider that a potential filament detected at pixel $i$ is significant if
\begin{equation}
    \label{eq:chi_r}
    (\chi_{\rm r}^{\rm M})_i < (\chi_{\rm r})_{\rm th}\,,
\end{equation}

\noindent or, equivalently,

\begin{equation}
    \label{eq:criterion_significance}
    S_i>1 \, ,
\end{equation}

\noindent where

\begin{equation}
    \label{eq:significance}
    S_i = \frac{(\chi_{\rm r})_{\rm th}}{(\chi_{\rm r}^{\rm M})_i}\, 
\end{equation}

\noindent is defined as the significance of the detection.
The definition of the threshold $(\chi_{\rm r})_{\rm th}$ in Sect.~\ref{sec:montecarlo}
implies that there is a 5$\,\%$ chance of mistakenly rejecting an ideal filament that was actually significant.

\subsubsection{Description of the Monte-Carlo simulations}
\label{sec:montecarlo}

The purpose of the Monte-Carlo simulations (illustrated in Fig.~\ref{fig:binary_image_significance}) is to derive the threshold, $(\chi_{\rm r})_{\rm th}$, below which a potential filament can be considered significant against the background (see Eq.~\ref{eq:chi_r}). This threshold depends only on the bar width, $W_{\rm b}$, and it applies to the entire map A. At each iteration (top row), a pixel $j$ is drawn at random in map A, and a sub-region A$_j$ of size $L_{\rm b}$, centered on pixel $j$, is cut-out (leftmost panel of Fig.~\ref{fig:binary_image_significance}). Pixel $j$ must lie far enough from the border of map A to ensure that A$_j$ is entirely contained within A and that convolution with a 2D top-hat function of radius $R$ will be possible (as explained in Sect.~\ref{sec:binary_map}). A synthetic map A$^{\rm S}_j$ is then created (middle left panel) by superposing on the sub-region A$_j$ an ideal filament with uniform intensity $\I_0$, centered on pixel $j$ and oriented at random (over a uniform distribution in angle). For convenience, $\I_0$ is written in terms of the standard deviation of sub-region A$_j$, $\sigma_{{\rm A}_j}$, and the signal-to-noise ratio of the filament, $\SNRfil$: $\I_0 = \sigma_{{\rm A}_j} \ \SNRfil$. $\SNRfil$ is a free parameter, whose value is discussed in Sect.~\ref{sec:instructions_for_use}.

Applying {\tt FilDReaMS} to the synthetic map A$^{\rm S}_j$ (surrounded by a band of width $R$ from map A to allow convolution with a 2D top-hat function of radius $R$
\footnote{For each value of the bar width, $W_{\rm b}$, the value of $R$ is calculated once and for all for the entire map A (see Sect.~\ref{sec:binary_map}); it is not recalculated at each Monte-Carlo iteration.}) leads to a synthetic binary map C$^{\rm S}_j$ (middle right panel of Fig.~\ref{fig:binary_image_significance}) followed by a synthetic orientation function, $f^{\rm S}_j$ (blue curve in the right panel of Fig.~\ref{fig:binary_image_significance}). Adding a contrasted synthetic filament in A$^{\rm S}_j$ implies that all the pixels close to the filament are now part of a less contrasted background (with respect to the filament), and together form a thin dark region the surrounding filament in C$^{\rm S}_j$ (see Sect.~\ref{sec:binary_map}). The synthetic orientation function $f^{\rm S}_j$ can be compared to the corresponding ideal orientation function, $f^{\rm I}_j$, similar to the orientation function obtained for the same ideal filament superposed on an empty background (orange curve).
The associated $(\chi_{\rm r}^{\rm S})_j$ is then derived from Eq.~\ref{eq:delta} with $f^{\rm M}$ replaced by $f^{\rm S}$. Clearly, $(\chi_{\rm r}^{\rm S})_j$ quantifies the impact of the background on the ideal filament by comparing the orientation functions $f^{\rm I}$ and $f^{\rm S}$ obtained from maps without (map A$_i$ in Fig.~\ref{fig:Ideal_hist_window}, for example) and with (map A$^{\rm S}_j$ in Fig.~\ref{fig:binary_image_significance}) background, respectively.

This Monte-Carlo iteration is repeated ten thousand times (each time with a new random region A$_j$ and a new random orientation of the ideal filament). The resulting Monte-Carlo distribution $D$ of $\chi_{\rm r}^{\rm S}$ is shown in the bottom panel of Fig.~\ref{fig:binary_image_significance}. The threshold $(\chi_{\rm r})_{\rm th}$ is chosen to be the 95-th percentile of the distribution of $\chi_{\rm r}^{\rm S}$.

\subsection{Reconstruction of physical filaments}
\label{sec:filament_visualisation}

So far, we have used a model bar with given width $W_{\rm b}$, and we have applied {\tt FilDReaMS} to a given pixel $i$ of the initial map A.
More specifically, we have detected potential bar-like filaments centered at pixel $i$ (Sect.~\ref{sec:orientation_angle}) and we have retained the significant filaments, similar to filaments with significance $S_i > 1$ (Sect.~\ref{sec:significance}).

\begin{figure}
    \centering
    \includegraphics[width=\columnwidth]{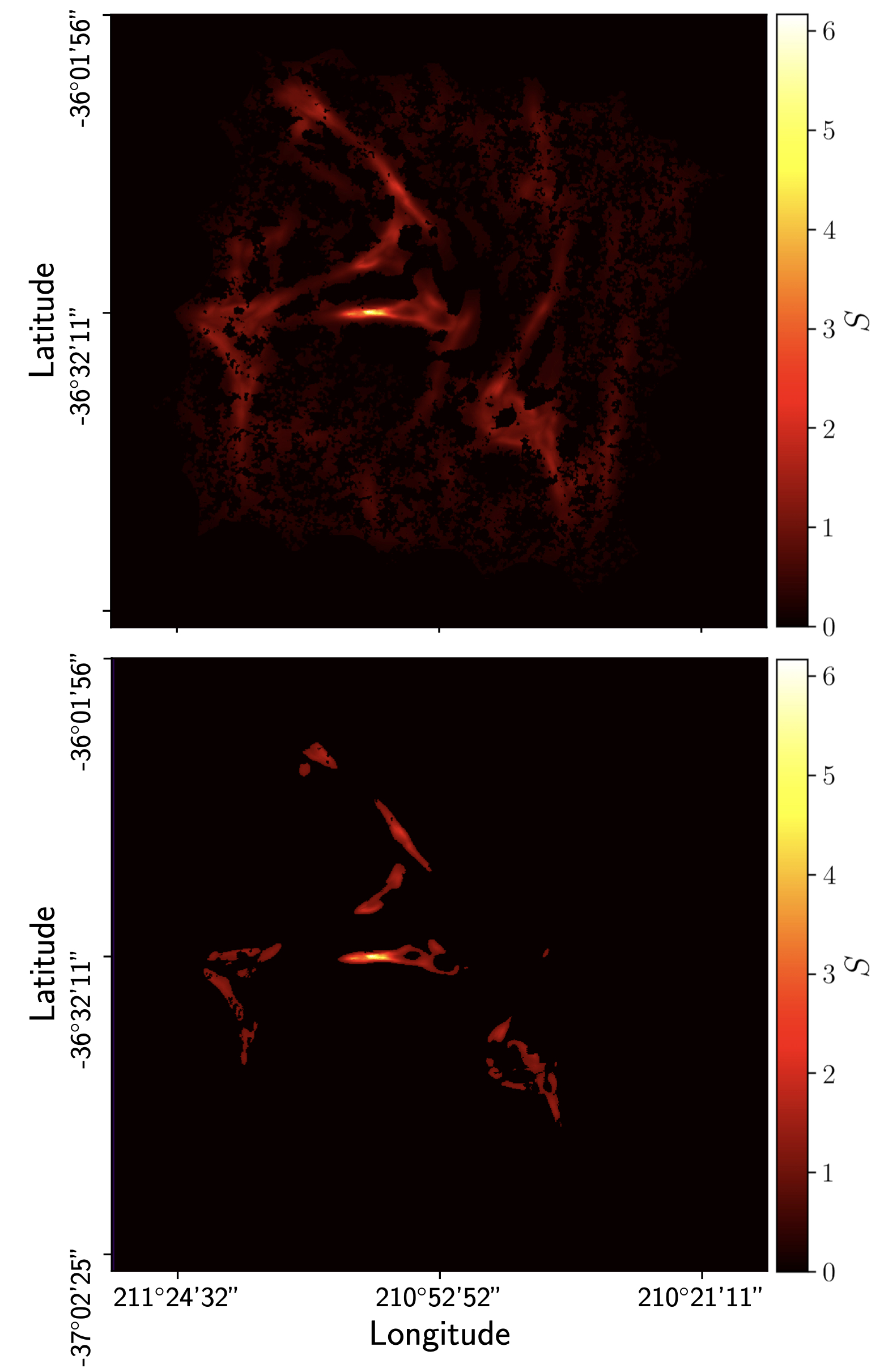}
    \caption{Maps of the significance, $S$ (Eq.~\ref{eq:significance}), of the G210 {\it Herschel} field for a bar width $W_{\rm b}=14\,{\rm px}$: ({\bf top}) for all the pixels of the binary map C and ({\bf bottom}) for the pixels that are the centers of significant filaments ($S>1$).}
    \label{fig:significance_maps}
\end{figure}

To detect all the bar-like filaments of bar width $W_{\rm b}$ in map A, we repeat the procedure for all the pixels
$i$ of map C that are more distant than $L_{\rm b}/2$ from the border (see beginning of Sect.~\ref{sec:histogram_of_orientation}). The significance at all the pixels of C (for $W_{\rm b}=14\,{\rm px}$) is shown in the top panel of Fig.~\ref{fig:significance_maps}, while the significance at the centers of significant filaments ($S>1$) are shown in the bottom panel. We can see that most pixels of map C are not significant ($S<1$). The most significant filaments have $S\simeq6$.

\begin{figure*}
    \centering
    \includegraphics[width=\textwidth]{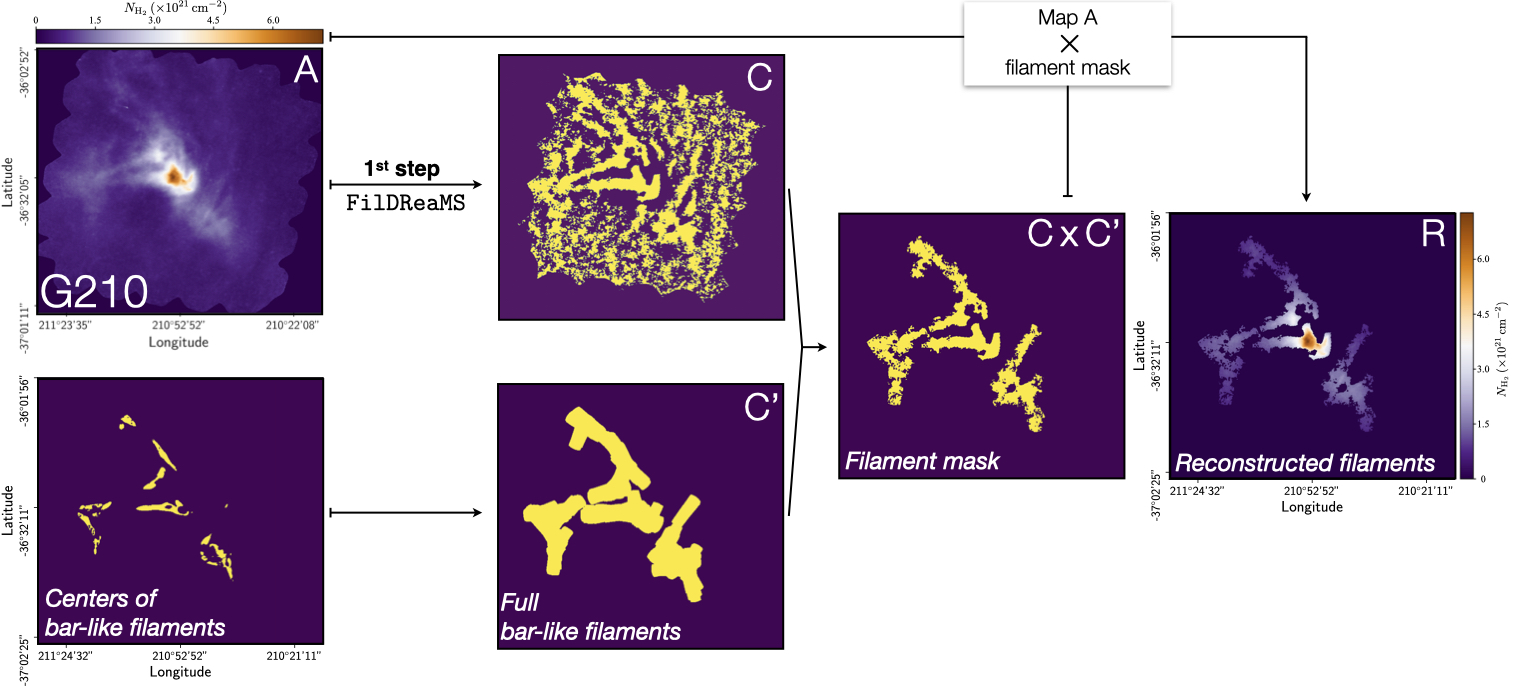}
    \caption{Illustration of the method used to reconstruct physical filaments of bar width ($W_{\rm b}=14\,{\rm px}$).
    {\bf Top left}: Initial map A of the G210 field.
    {\bf Top middle}: Binary map C from the first step of {\tt FilDReaMS} applied to map A (same pixels as in the top panel of Fig.~\ref{fig:significance_maps}).
    {\bf Bottom left}: Pixels at which at least one significant bar-like filament of bar width $W_{\rm b}$ is detected (same pixels as in the bottom panel of Fig.~\ref{fig:significance_maps}).
    {\bf Bottom middle}: Binary map C' in which the model bars associated with all the significant bar-like filaments are set to 1 and the background is set to 0.
    {\bf Right}: Filament mask obtained by multiplying C by C'.
    {\bf Rightmost}: Reconstructed physical filaments obtained by applying the filament mask to the initial map A.}
    \label{fig:filament_reconstruction}
\end{figure*}

By combining all the significant bar-like filaments, we can then reconstruct the true shape and the intensity of physical filaments, as illustrated in Fig.~\ref{fig:filament_reconstruction}.
We first produce a binary map C' in which the model bars associated with all the significant filaments have their pixels set to 1, while all the other pixels are set to 0. We then create a filament mask by multiplying the binary map C introduced in Sect.~\ref{sec:binary_map} by the new binary map C', and we apply this mask to the initial map A, by computing a simple product of both images.
The resulting map R (rightmost panel of Fig.~\ref{fig:filament_reconstruction}) reveals the network of physical filaments of bar width $W_{\rm b}$.

At this point, a filament orientation angle (denoted by $(\psi_{\rm f})_i$ at pixel $i$) is defined only at pixels at which one or more significant bar-like filaments are found (bottom-left panel of Fig.~\ref{fig:filament_reconstruction}).
We now assign a filament orientation angle (denoted by $(\psi_{\rm f}^\star)_{i'}$ at pixel $i'$) to all non-zero pixels of map R
(rightmost panel of Fig.~\ref{fig:filament_reconstruction}).
For each pixel $i'$, we consider all the significant filaments whose associated model bars pass through $i'$, and we define $(\psi_{\rm f}^\star)_{i'}$ as the orientation angle of the most significant filament, similar to the filament with the highest significance, $S_i$ (Eq.~\ref{eq:significance}).
It is important to realize that for pixels $i'$ at which $(\psi_{\rm f})_{i'}$ is defined, $(\psi_{\rm f}^\star)_{i'}$ generally differs slightly from $(\psi_{\rm f})_{i'}$, because the most significant bar-like filament passing through $i'$ is generally not the filament centered on $i'$, but a filament centered on a neighboring pixel $i$. Also keep in mind that $(\psi_{\rm f}^\star)_{i'}$ is a function of the bar width, $W_{\rm b}$.

\subsection{Derivation of most significant bar widths and most prevalent bar widths}
\label{sec:signif_bar_width}

Now that we have laid out the entire procedure for a given bar width, $W_{\rm b}$ (Sects~\ref{sec:binary_map} to \ref{sec:filament_visualisation}), we iterate over a range of values of $W_{\rm b}$
(see Sect.~\ref{sec:instructions_for_use} for a recommendation on the  optimal range). This enables us to assign a "most significant bar width", $(W_{\rm b}^\star)_{i'}$, to every non-zero pixel $i'$ of the different reconstructed maps R$(W_{\rm b})$.
Namely, for every pixel $i'$, we consider all the significant filaments whose associated model bars pass through $i'$, and we define $(W_{\rm b}^\star)_{i'}$ as the bar width of the most significant filament.

We can then construct a histogram of $W_{\rm b}^\star$ over all pixels, namely,
the number of pixels, $N_{\rm pix}$, whose most significant bar width is equal to $W_{\rm b}^\star$. 
In practice, to make it easier to compare different maps, we propose to work with a normalized histogram, where $N_{\rm pix}$ is divided by the total number of pixels in the map, $N_{\rm map}$.
If the histogram exhibits clear peaks (i.e., local maxima) the corresponding bar widths are denoted by $W_{\rm b}^{\star{\rm peak}}$ and referred to as "most prevalent bar widths".
Examples of histograms $(N_{\rm pix}/N_{\rm map})$ versus $W_{\rm b}^\star$ are shown in Figs.~\ref{fig:power_spectrum} and \ref{fig:plummer_power_spectrum}.

\begin{table*}
\caption{Free parameters of {\tt FilDReaMS}, with their definitions (second column) and the corresponding characteristics of the filaments to be detected by {\tt FilDReaMS} (third column).}
\centering
\begin{threeparttable}
\newcolumntype{M}[1]{%
 >{\vbox to 2ex\bgroup\vfill}%
 p{#1}%
 <{\egroup}} 
\begin{tabular}{m{2.0cm} | M{6.5cm} | M{6.5cm}}
\midrule\midrule
$W_{\rm b}$ & Width of the {\tt FilDReaMS} model bar & Width of the detected filaments\\
\cmidrule(l  r ){1-3}
$r_{\rm b}=L_{\rm b}$/$W_{\rm b}$ & Aspect ratio of the {\tt FilDReaMS} model bar & Minimum elongation of the detected filaments\\
\cmidrule(l  r ){1-3}
$\SNRfil$ & Signal-to-noise ratio of the ideal filament in the Monte-Carlo simulations & Minimum contrast of the detected filaments\\
\midrule\midrule
\end{tabular}
\end{threeparttable}
\label{tab:FilDReaMS_parameters}
\end{table*}

\subsection{Instructions for use}
\label{sec:instructions_for_use}

We now discuss the optimal values, or ranges of values, of the three important free parameters of {\tt FilDReaMS} (see Table~\ref{tab:FilDReaMS_parameters}): the width of the model bar, $W_{\rm b}$, the aspect ratio of the model bar, $r_{\rm b} = L_{\rm b} / W_{\rm b}$, and the signal-to-noise ratio of the ideal filament in the Monte-Carlo simulations, $\SNRfil$. 

The aspect ratio of the model bar, $r_{\rm b}$, sets an approximate lower limit to the aspect ratio of elongated structures that can be detected with {\tt FilDReaMS}. Here, we consider that an elongated structure can be qualified as a filament if its aspect ratio is at least 3. Accordingly, we recommend to adopt $r_{\rm b}=3$. This value was also used by \citet{Panopoulou_2014,Arzoumanian_2019}.

The signal-to-noise ratio of the ideal filament in the Monte-Carlo simulations, $\SNRfil$, sets an approximate lower limit to the intensity $\I$ (in a broad sense) of filaments that have a high probability of being detected with {\tt FilDReaMS}. If $\SNRfil$ is too low, some of the small-scale noise structures might be mistaken for physical filaments. On the other hand, if $\SNRfil$ is too high, some physical filaments might escape detection. As a compromise, we recommend to adopt $\SNRfil=3$.

The width of the model bar, $W_{\rm b}$, is only constrained by the pixel size at the low end and by the size of the initial map A at the high end. 
For the lower limit, a good choice is typically $(W_{\rm b})_{\rm min} = 5\,{\rm px}$; this choice is particularly relevant in the case of the {\it Herschel} fields, where the beam size equals three times the pixel size.
For the upper limit, we take $(W_{\rm b})_{\rm max} = (L_{\rm b})_{\rm max} / r_{\rm b}$, with $(L_{\rm b})_{\rm max}$ equal to one-third the size of map A;
in the {\it Herschel} G210 field, $(W_{\rm b})_{\rm max} = 27\,{\rm px}$.

\section{Validation}
\label{sec:Validation}

{\tt FilDReaMS} is in principle able to detect filaments with different sizes and orientations in a given map of intensity $\I$. Before applying {\tt FilDReaMS} to real scientific data, we need to validate the method and estimate its reliability.

In the following, we explore the impact of the filament profile, the noise level and the aspect ratio on the bar width and orientation estimates. We finally investigate the case with superposition of filaments with variable intensities, in combination with the above parameters.

We first describe in Sect.~\ref{sec:simulations} the set of simulations used to performed these analyses. We then test the ability of {\tt FilDReaMS} to recover the widths of the input filaments in Sect.~\ref{sec:fil_scales}, and their orientations in Sect.~\ref{sec:fil_orientations}.

\subsection{Set of Simulations}
\label{sec:simulations}

We create several series of simulated maps composed of synthetic filaments embedded in realistic environments, including noise and incoherent structures. To study the impact of the filament radial profile, we consider in these simulations two types of synthetic filaments, which, by default (i.e., unless explicitly stated otherwise), have the following characteristics:
\begin{itemize}
    \item Ideal filaments: They have the shape of the rectangular model bar, with width $W_{\rm b}$, length $L_{\rm b}$, and aspect ratio $r_{\rm b} = L_{\rm b} / W_{\rm b}$, and they have uniform intensity, $\I_0$. \\
    
    \item Plummer-type filaments: Their intensity can be described by a 2D Plummer-type profile with half-width $R_{\rm flat}$ and aspect ratio $r_{\rm b}$:
    
    \begin{equation}
    \label{eq:plummer}
        \I_{\rm P}(x, y) = \I_0\left[ 1 + \left( \frac{x}{R_{\rm flat}} \right)^2 + \left( \frac{y}{r_{\rm b} \, R_{\rm flat}} \right)^2 \right]^{-(p-1)/2}\,,
    \end{equation}
    
   \noindent where $x$ and $y$ are the coordinates across and along the long axis of the filament, $p$ is the Plummer power-law index, and $\I_0$ is the central intensity. We adopt $p=2.2$ \citep[median value obtained by][for a sample of 599 filaments including G300]{Arzoumanian_2019}.
\end{itemize}

The realistic environment is simulated as a combination of white and power-law (Brownian) noises, with r.m.s. intensity $\sigma_{\mathcal{W}}$ and $\sigma_{\mathcal{B}}$, respectively. The white noise is meant to represent instrumental noise, which was estimated to be $\leq 7\%$ of the intensity signal for {\it Herschel} SPIRE data \citep[][]{Juvela_GCCIII_2012}, while the Brownian noise (with a power-law index of $-2$) is meant to represent astrophysical fluctuations \citep[e.g.][]{Miville-Deschenes_2003_brownian}. Both types of noise have different realisations in each map. In the following, we consider two different configurations denoted as 'default' and 'high' levels of noise. In the first case, we chose $\sigma_{\mathcal{W}} = 0.05\,\I_0$ and $\sigma_{\mathcal{B}} = 0.3\,\I_0$, while in the second case we chose $\sigma_{\mathcal{W}} = 0.1\,\I_0$ and $\sigma_{\mathcal{B}} = \I_0$.

We perform four different sets of simulations: \begin{itemize}[noitemsep,topsep=0pt]
    \item \textbf{SimSet 1}: $r_{\rm b}$=3 and default noise level.
    \item \textbf{SimSet 2}: $r_{\rm b}$=3 and high noise level.
    \item \textbf{SimSet 3}: $r_{\rm b}$=10 and default noise level.
    \item \textbf{SimSet 4}: Overlapping filaments with variable aspect ratios and intensities, with default noise level.
\end{itemize}

For the first three sets of simulations, it consists of the co-addition of a map of synthetic filaments with realisations of realistic noise. Each original map is made of 91 non-overlapping synthetic filaments of common width, $W_{\rm b}$, and aspect ratio, $r_{\rm b}$, but different orientations spanning linearly from 0$^{\circ}$ to +90$^{\circ}$ in 1$^{\circ}$ steps. A series of 16 maps are thus created for each value of the width, $W_{\rm b}$ ranging from 5 to 20\,px (in steps of 1\,px), on which 2000 realisations of realistic noise are co-added.

For the last set of simulations, SimSet~4, it consists of a single series of 100 maps containing 64 synthetic filaments: 48 filaments with $W_{\rm b}=9\,{\rm px}$ and 16 filaments with $W_{\rm b}=15\,{\rm px}$. Both groups of filaments cover approximately the same total surface area. Each filament is randomly assigned one out of 8 possible values of the filament aspect ratio, $r_{\rm b}$, in the linear range $[3,10]$ and (independently) one out of 8 possible values of the filament intensity, $\I_0$, in the logarithmic range $[0.2,5]$. Each value of $r_{\rm b}$ and each value of $\I_0$ are assigned to exactly eight filaments. The filaments are located at random positions, they have random orientations, and they can possibly overlap. For each map realisation, we co-add a noise realisation using the default configuration. 

All sets of simulations are finally repeated for ideal and Plummer-type filaments, and their simulation parameters are displayed in Table.~\ref{tab:simulation_sets}. In the case of SimSet~1, we also produce two other sets of simulations using values of the Plummer-type filaments with power-law index, $p=1.2$ and 4. The value $p = 4$ corresponds to an isolated filament in isothermal, non-magnetic hydrostatic balance \citep{Ostriker_1964}; this value was also used by \cite{Suri_2019} to fit filament profiles. The value $p = 1.2$ is the lower limit explored by \cite{Arzoumanian_2019} in their study. The larger and smaller values yield more and less contrasted filaments, respectively.

\begin{table*}
\centering 
\caption{Description of the parameters of the various sets of simulations performed to test the robustness of the width and orientation estimates recovery by {\tt FilDReaMS}.\\}
\begin{threeparttable}
\begin{tabular}{c | cccccc }
\toprule\toprule
\multirow{3}{*}{\begin{tabular}{c} \textbf{\large SimSet} \\ \textbf{ID} \\ \end{tabular}}
 & \multicolumn{6}{c}{Simulation parameters}\\ 
\cline{2-7}
& width & orientation & aspect ratio & central intensity & overlap & noise level \\ 
& $[\rm{px}]$ &  & $r_{\rm{b}}$ & $\I_0$ & &  $(\sigma_{\mathcal{W}}, \sigma_{\mathcal{B}}) \% \times \I_0$ \\
\hline
 \textbf{SimSet 1}  & $[5,20]$ &  $[0^{\circ},90^{\circ}]$\tnote{a} & 3 & 1 & no & $(5,30)$ \\
 \textbf{SimSet 2}  & $[5,20]$ &  $[0^{\circ},90^{\circ}]$\tnote{a} & 3 & 1 & no & $(10,100)$ \\
 \textbf{SimSet 3}  & $[5,20]$ &  $[0^{\circ},90^{\circ}]$\tnote{a} & 10 & 1 & no & $(5,30)$ \\
 \textbf{SimSet 4}  & $9; 15$ &  $[0^{\circ},90^{\circ}]$\tnote{b}  & $[3,10]$\tnote{c} & $[0.2,5]$\tnote{d} & yes &  $(5,30)$
\\
\bottomrule\bottomrule
\end{tabular}
  \begin{tablenotes}
    \item {\bf Notes.} The 'bar width' (in pixel), 'orientation', 'aspect ratio' ($r_{\rm{b}}$), and 'central intensity' ($\I_0$) provide the ranges of those parameters used to build the simulated bars, which are specified to present 'overlap' or not ; the 'noise level' is described with two components, a level of white noise ($\sigma_{\mathcal{W}}$) and Brownian noise ($\sigma_{\mathcal{B}}$).
    \item[a] All values are considered within the interval with a step of $1^{\circ}$.
    \item[b] Randomly picked in the considered range.
    \item[c] Randomly assigned to one out 8 possible values in the interval, following linear steps of $1\,\rm{px}$.
    \item[d] Randomly assigned to one out 8 possible values in the interval, distributed in logarithmic scale.
  \end{tablenotes}
\end{threeparttable}
\label{tab:simulation_sets}
\end{table*}

\subsection{Filament bar widths}
\label{sec:fil_scales}

In the following, we construct the histogram $(N_{\rm pix}/N_{\rm map})$ versus $W_{\rm b}^\star$ of every map (as explained in Sect.~\ref{sec:signif_bar_width}), and we average over the 2000 noise realisations for the first three sets of simulations (SimSet~1, 2 and 3) and over the 100 noise realisations for the fourth set of simulations (SimSet~4). For this analysis, we can identify a most prevalent bar width $W_{\rm b}^{\star \rm peak}$,  averaged over the multiple noise realisations, and its uncertainty (or precision), equivalent to the standard deviation of the histogram. The most prevalent bar width will then be compared with the input width of the simulation. 

\subsubsection{Impact of the noise level}
\label{sec:fil_scales_noise}

For this study we focus on the first two sets of simulations, SimSet~1 \& 2.

\begin{figure}
    \centering
    \includegraphics[width=\columnwidth]{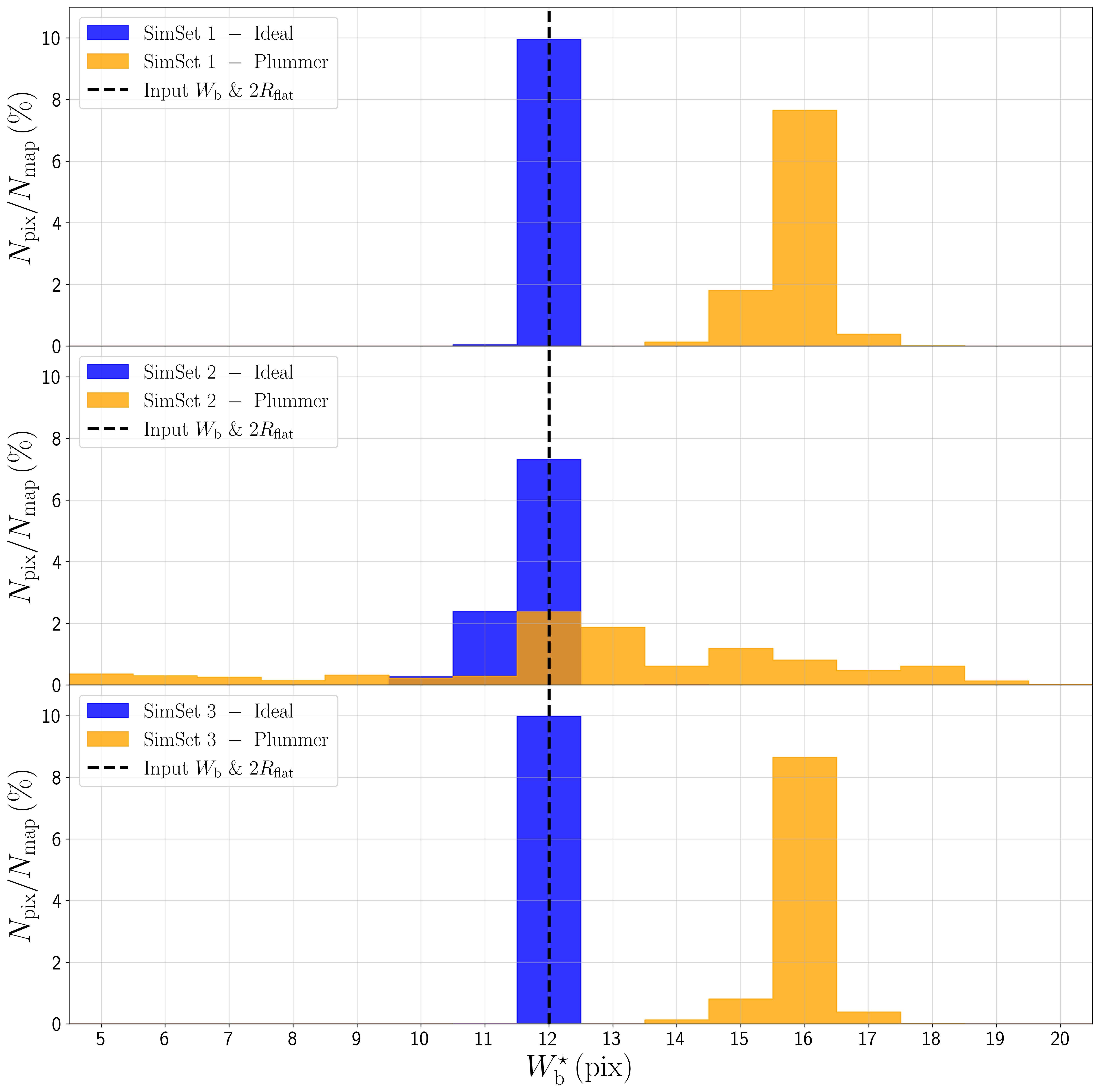}
    \caption{Average histograms $(N_{\rm pix}/N_{\rm map})$ versus $W_{\rm b}^\star$ obtained for ideal (blue) and Plummer-type (orange, with power-law index $p=2.2$) filaments with bar width $W_{\rm b}= 12\,{\rm px}$ and Plummer half-width $R_{\rm flat}= 12\,{\rm px}$, respectively, in three different sets of simulations, SimSet~1 ({\bf top}), SimSet~2 ({\bf middle}) and SimSet~3 ({\bf bottom}).
    }
  \label{fig:test_histo}
\end{figure}

In the case of ideal filaments, we find for each input value of $W_{\rm b}$ that the average histogram is dominated by a well-defined peak at $W_{\rm b}^{\star{\rm peak}} = W_{\rm b}$, which means that {\tt FilDReaMS} exhibits a very good accuracy by easily recovering the right width of the input filament, even in the presence of high noise. However, the peak does not stand out as clearly at the high noise level (SimSet~2) as at the default noise level (SimSet~1). The precision of the width estimate is very stable and is only moved from $0.01\,{\rm px}$ to $0.4\,{\rm px}$ with increasing noise level (see Table.~\ref{tab:uncertainties}).
This is illustrated for $W_{\rm b}= 12\,{\rm px}$ in Fig.~\ref{fig:test_histo}, where the histogram at the high noise level (middle panel, blue) can be compared to the histogram at the default noise level (top panel, blue).

We show the case of the Plummer-type filaments for the same sets of simulations in the top and middle panels of Fig.~\ref{fig:test_histo} (orange histograms), with $R_{\rm flat}= 12\,{\rm px}$. In the case of default noise level (top panel), the histogram of $W_{\rm b}^{\star}$ is clearly shifted compared to the input value and more spread than in the case of the ideal filament. This shift is expected and will depend on the power-law of the Plummer profile, as illustrated in Sect.~\ref{sec:correspondance}. The larger spread can be explained here by the larger impact of the noise on the tails of the Plummer profile.
In addition, the central intensity $\I_0$ is only reached along the crest of the Plummer-type filament. The integral of the transverse profile of a Plummer-type filament over a bar width $W_{\rm b} = 2R_{\rm flat}$ represents $75\,\%$ of the integration over the corresponding ideal filament (when computed with the default Plummer power-law index equal to 2.2). Therefore, there will be a slightly higher degradation from the noise as one moves away from the crest. 
Despite the shift observed in the case of the Plummer-type filament with default noise, we still obtain a precision of $0.4\,{\rm px}$. However, the histograms $(N_{\rm pix}/N_{\rm map})$ versus $W_{\rm b}^\star$ of Plummer-type filaments may exhibit one or two spurious (lower) peaks.

The impact of increasing the noise level is much more pronounced than in the ideal filament case (see middle panel), leading to much larger uncertainties, $2.7\,{\rm px}$, stable over the whole range of input width (see Table.~\ref{tab:uncertainties}). The increase of noise level increases the frequency and the strength of the spurious peaks. The observed shift between the widths $2R_{\rm flat}$ and $W_{\rm b}^{\star, \rm peak}$ can be empirically predicted, as we will discuss in Sect.\ref{sec:correspondance}.

\subsubsection{Impact of the filament aspect ratio}
\label{sec:fil_scales_aspectratio}

For this analysis we use the two sets of simulations, SimSet~1 \& 3, between which the aspect ratio has been changed from 3 to 10. The peak becomes increasingly dominant as $r_{\rm b}$ increases, leading to a gain of a factor of 2 to 5 of the precision, for Plummer and ideal cases respectively (see Table.~\ref{tab:uncertainties}). This is in accordance with our expectation that more elongated filaments are easier to recognize. In the case of Plummer-type filaments, the shift between $2R_{\rm flat}$ and $W_{\rm b}^{\star, \rm peak}$ appears to be unchanged. This is illustrated in Fig.~\ref{fig:test_histo}, where the histograms obtained for $r_{\rm b}= 10$ (bottom panel) can be compared to the histograms obtained for $r_{\rm b}= 3$ (top panel).

\subsubsection{Plummer-to-ideal width relation}
\label{sec:correspondance}

\begin{figure}
    \centering
    \includegraphics[width=0.85\columnwidth]{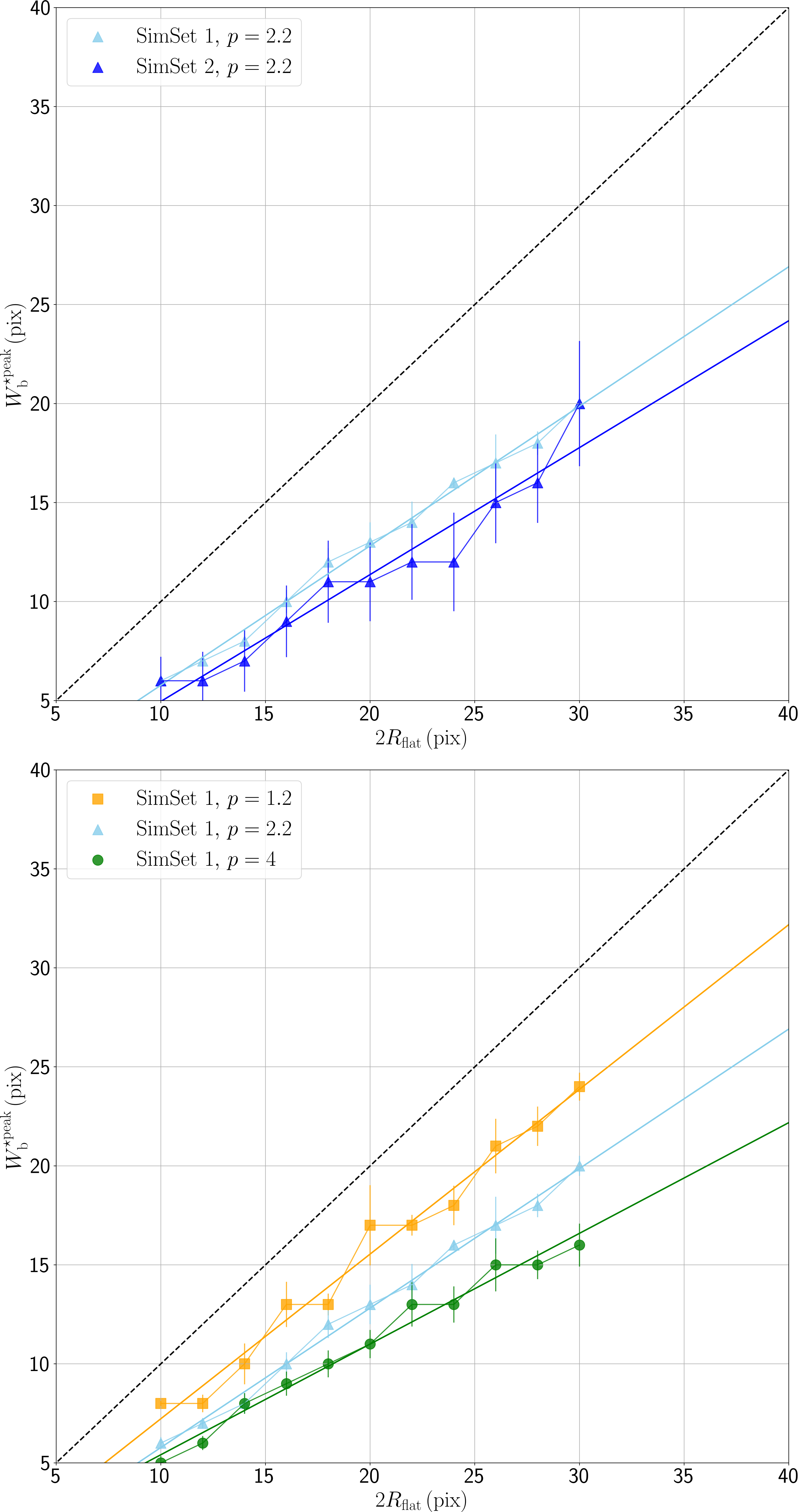}
    \caption{Average relation between the width $2R_{\rm flat}$ of an input Plummer-type filament and the most prevalent bar width $W_{\rm b}^{\star{\rm peak}}$ derived by {\tt FilDReaMS}, for two different noise levels ({\bf top}) and for three different values of the Plummer index, $p$ ({\bf bottom}).
    The solid lines are linear fits to the relations $W_{\rm b}^{\star{\rm peak}}$ versus $2R_{\rm flat}$ of the same colors, and the dashed line is the straight line $W_{\rm b}^{\star{\rm peak}} = 2R_{\rm flat}$.}
    \label{fig:scale_plum_rect}
\end{figure}

The sets of simulations with Plummer-type filaments allow us to derive the empirical relation between $2R_{\rm flat}$ and $W_{\rm b}^{\star{\rm peak}}$ for different noise levels and also for different values of the Plummer index, $p$.
The curves linking the input $2R_{\rm flat}$ to the output $W_{\rm b}^{\star{\rm peak}}$ are plotted in the top panel of Fig.~\ref{fig:scale_plum_rect}, for the default and high noise levels, SimSet~1 (blue triangles) and SimSet~2 (dark blue triangles) respectively. Also plotted are the line $W_{\rm b}^{\star{\rm peak}} = 2R_{\rm flat}$ (dashed line), the linear fits to the two curves $W_{\rm b}^{\star{\rm peak}}$ versus $2R_{\rm flat}$ (solid lines in the corresponding colors), as well as the standard deviations computed on each average histogram $(N_{\rm pix}/N_{\rm map})$ versus $W_{\rm b}^\star$ (error bars on each symbol in the corresponding colors).
For future reference, the parameters $a$ and $b$ of the linear fits $W_{\rm b}^{\rm \star peak} = a \, (2R_{\rm flat}) + b$ are given in Table~\ref{tab:linear_fits}.

The error in the fit for SimSet~1 can be roughly estimated at $\lesssim 1\,{\rm px}$ over the range $R_{\rm flat} = [5, 20]\,{\rm px}$, from a comparison with linear fit constrained to pass through the origin.
This error adds to the statistical uncertainty $\simeq 0.5\,{\rm px}$ associated with the dispersion of the measured points about the fits and to the standard deviations (see Table.~\ref{tab:uncertainties}).
In the case of high noise level (SimSet~2), $a$ slightly decreases with increasing noise, whereas $b$ takes on larger values. 
This, combined with the high statistical uncertainty and standard deviations (error bars), indicates that {\tt FilDReaMS} is not well suited to recover the width of input Plummer-type filaments in the case of high noise level.

\begin{table*}
\centering 
\caption{Parameters $a$ and $b$ of the linear fits $W_{\rm b}^{\rm \star peak} = a \, (2R_{\rm flat}) + b$ to the curves plotted in Fig.~\ref{fig:scale_plum_rect}, in the case of two noise-levels (SimSet1 and SimSet2) and three Plummer power-law index $p=1.2, 2.2,$ and $4$.}
\begin{tabular}{c | cccc }
\toprule\toprule
\multirow{2}{*}{\begin{tabular}{c} \textbf{\large Linear fit} \\ \textbf{\large parameters} \\\end{tabular}}
& \multicolumn{4}{c}{SimSet ID \& Plummer power-law index $p$}\\ 
\cline{2-5}
& \textbf{SimSet 1 ($p=2.2$)} & \textbf{SimSet 2 ($p=2.2$)} & \textbf{SimSet 1 ($p=4$)} & \textbf{SimSet 1 ($p=1.2$)} \\ 
\hline
 Slope $a$  & 0.70 & 0.64 & 0.56 & 0.83 \\
 Origin $b$  & -1.27 & -1.45 & -0.18 & -1.09 \\
\bottomrule\bottomrule
\end{tabular}
\label{tab:linear_fits}
\end{table*}

For completeness, we also look at the impact of the Plummer index on the curves $W_{\rm b}^{\star{\rm peak}}$ versus $2R_{\rm flat}$.
Plotted in the bottom panel of Fig.~\ref{fig:scale_plum_rect} are the curves obtained for $p = 1.2$ (orange squares), $p = 2.2$ (blue triangles), and $p = 4$ (green circles) with their respective standard deviations (error bars in the corresponding colors).
The linear fits to the three curves $W_{\rm b}^{\star{\rm peak}}$ versus $2R_{\rm flat}$ are again shown in solid lines, and their parameters given in Table~\ref{tab:linear_fits}.
The parameter $a$ increases with decreasing $p$. The increase in slope can be understood by noting that, for a given $R_{\rm flat}$, a smaller $p$ yields a more spread-out Plummer-type filament (see Eq.~\ref{eq:plummer}), which in turn results in a detection at larger $W_{\rm b}^\star$.
As for $p=2.2$, the error in the fits for the other two values of $p$ can be roughly estimated at $\lesssim 1\,{\rm px}$ over the range $R_{\rm flat} = [5, 20]\,{\rm px}$.

Finally, the curves $W_{\rm b}^{\star{\rm peak}}$ versus $2R_{\rm flat}$ obtained for increasing values of $r_{\rm b}$ are identical, with the standard deviations (error bars in Fig.~\ref{fig:scale_plum_rect}) decreasing down to half those obtained for $r_{\rm b}= 3$. From this we conclude that the empirical relation between $2R_{\rm flat}$ and $W_{\rm b}^{\star{\rm peak}}$ derived in SimSet~3 remains valid independently of $r_{\rm b}$.

\subsubsection{Overlapping filaments of different properties}
\label{sec:fil_scales_overlap}

To study reliability of width estimate in a more realistic case, we focus now on the fourth set of simulations, SimSet~4, for ideal and Plummer-type filaments.

\begin{figure}
    \centering
    \includegraphics[width=\columnwidth]{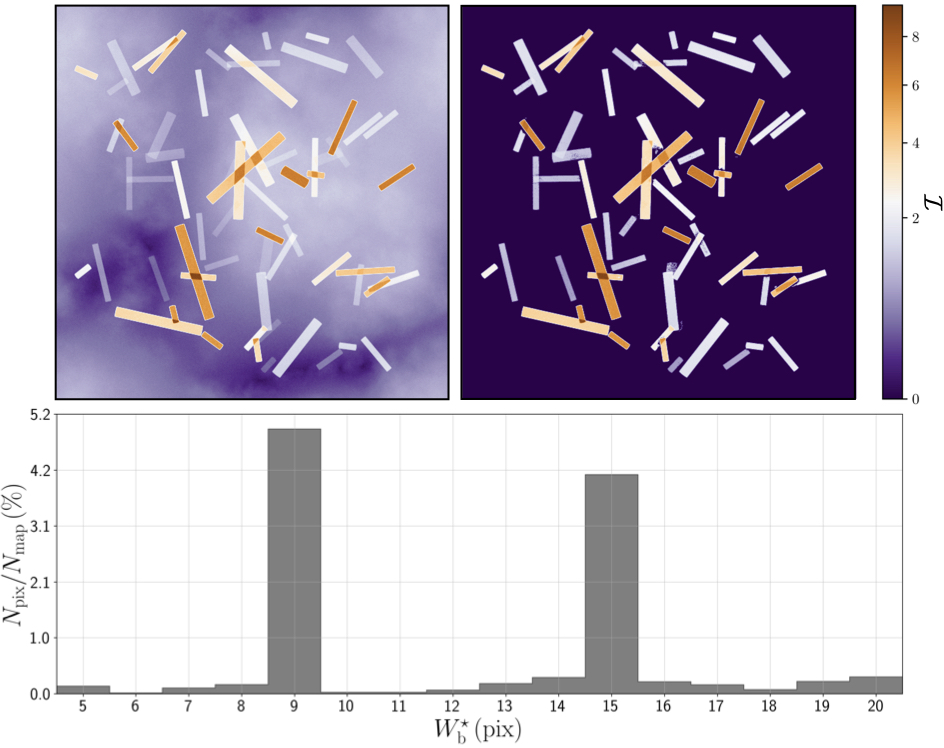}
    \caption{Visualisation of the set of simulations SimSet~4 for ideal filaments, showing how well {\tt FilDReaMS} recovers the widths $W_{\rm b}$ of overlapping filaments with different widths, aspect ratios, and intensities.
    {\bf Top left}: One of the 100 test maps, containing 64 ideal filaments, with width $W_{\rm b}=9\,{\rm px}$ or $15\,{\rm px}$, aspect ratio in the range $[3,10]$, and intensity in the range $[0.2,5]$.
    {\bf Top right}: Corresponding map of all the filaments detected and reconstructed by {\tt FilDReaMS}.
    {\bf Bottom}: Number of pixels, $N_{\rm pix}$, whose most significant bar width is $W_{\rm b}^{\star}$, normalized to the number of pixels in the map, $N_{\rm map}$, as a function of $W_{\rm b}^{\star}$.}
    \label{fig:power_spectrum}
\end{figure}

In the case of ideal filaments, {\tt FilDReaMS} is able to detect and reconstruct all of them (top right panel of Fig.~\ref{fig:power_spectrum}).
The histogram $(N_{\rm pix}/N_{\rm map})$ versus $W_{\rm b}^\star$, averaged over the 100 maps, has two pronounced peaks at $W_{\rm b}^{\star{\rm peak}}=9\,{\rm px}$ and $W_{\rm b}^{\star{\rm peak}}=15\,{\rm px}$ (bottom panel of Fig.~\ref{fig:power_spectrum}).
Hence, {\tt FilDReaMS} recovers again the widths of the input ideal filaments.
The reason why the histogram also contains non-negligible values away from the peaks is because the superposition of two input filaments can either lead to a structure that will be identified as a thicker filament or conversely cause the broader filament to effectively hide part of the narrower filament.

\begin{figure}
    \centering
    \includegraphics[width=\columnwidth]{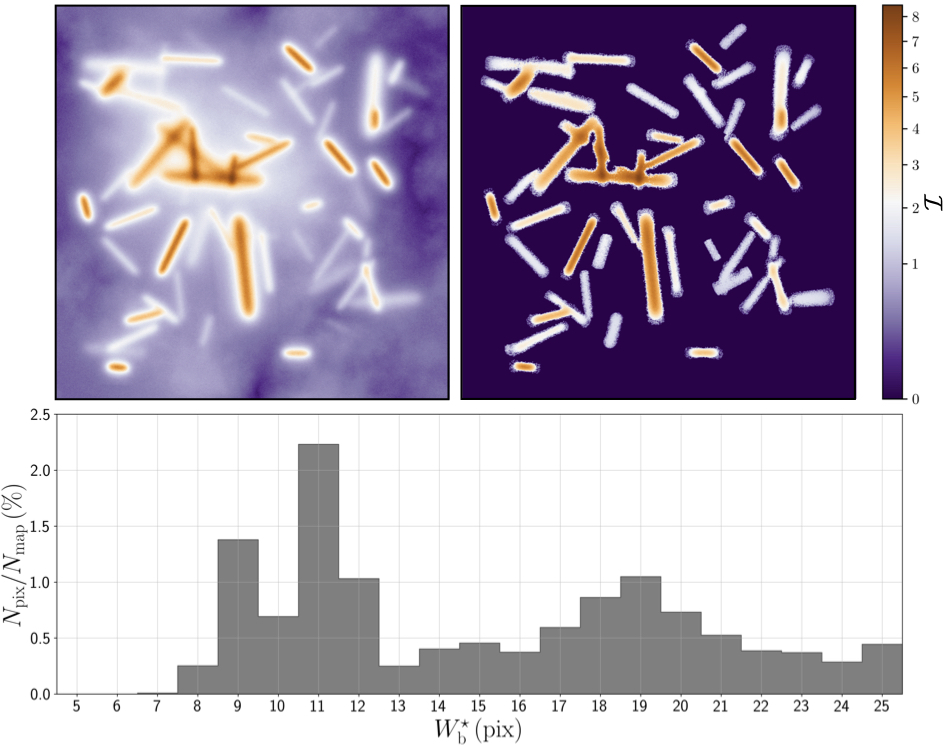}
    \caption{Same as Fig.~\ref{fig:power_spectrum}, with the input ideal filaments of width $W_{\rm b}$ replaced by Plummer-type filaments of width $2R_{\rm flat}$.}
    \label{fig:plummer_power_spectrum}
\end{figure}

In the case of Plummer-type filaments, {\tt FilDReaMS} is also able to detect and reconstruct all of them (top right panel of Fig.~\ref{fig:plummer_power_spectrum}).
The histogram $(N_{\rm pix}/N_{\rm map})$ versus $W_{\rm b}^\star$, averaged over the 100 test maps, has three peaks at $W_{\rm b}^{\star{\rm peak}}=9, 11\, \rm{and} \, 19\,{\rm px}$ which present a larger spread compared to the ideal case (bottom panel of Fig.~\ref{fig:plummer_power_spectrum}). The two peaks at $W_{\rm b}^{\star{\rm peak}}=9\, \rm{and}\, 11\,{\rm px}$ are associated with the same input filament width ($R_{\rm{flat}}=9\,\rm{px}$). This can be explained by the log-scale distribution of the central intensity of the filaments, $I_0$, which leads to two different noise regimes, with $1/3$ of the filaments in the high noise level, and $2/3$ in the default noise level. Following top panel of Fig.~\ref{fig:scale_plum_rect}, a $2R_{\rm{flat}}$ value of $18\,\rm{px}$ roughly corresponds to bar widths of $W_{\rm b}=9 \, \rm{and} \, 11\, \rm{px}$ in the high and default noise levels, respectively, which is consistent with the two observed peaks. The same kind of behaviour is not observed in this simulation for the population of larger filaments, since they are more affected by the impact of overlapping filaments, leading to a spread distribution masking this effect.

The corresponding values of $2R_{\rm flat}$ can be derived from the linear fits $W_{\rm b}^{\rm \star peak} = a \, (2R_{\rm flat}) + b$, with the values of $a$ and $b$ given in Table~\ref{tab:linear_fits}. We use the case $p=2.2$ and the high noise level for the first peak at $9\,rm{px}$ (as explained above), and the case $p=2.2$ with default noise level for the others. 
This leads to $2R_{\rm flat}=16.3\,{\rm px}$, $17.3\,{\rm px}$, and $28.5\,{\rm px}$, in good agreement with the input $R_{\rm flat}$ ($9\,{\rm px}$ and $15\,{\rm px}$). This agreement supports the general validity of the empirical relation between $2R_{\rm flat}$ and $W_{\rm b}^{\star{\rm peak}}$ obtained in Sect.~\ref{sec:correspondance}. As in the case of ideal filaments, the fact that the histogram contains more than the two expected peaks except the second peak at $9\,{\rm px}$ is not an indication that {\tt FilDReaMS} performs poorly, but rather a consequence of the subjective way of identifying filaments.

\subsection{Filament orientations}
\label{sec:fil_orientations}

Again, to study the impact of noise level and aspect ratio on the reliability of the filament orientation estimates, we use the first three sets of simulations SimSet~1, 2 and 3, for ideal and Plummer-type filaments. Hence, for each series, the derived orientation angles, $\psi_{\rm f}^{\star}$, are compared to the input orientations, $\psi_{\rm f}$ over the 2000 noise realisations of each set of simulation. This allows to estimate both the bias and the standard deviation $\sigma_{\psi}$ of $\psi_{\rm f}^{\star}$.

We first observe that the orientations of the reconstructed filaments don't show any deviation in average from the input orientation angles, and don't exhibit any specific dependency with the input value of the orientation angle. 
We also observe from these simulations that $\sigma_{\psi}$ decreases with the filament width, so that the results obtained with $W_{\rm b}$ and $R_{\rm flat} = 5\,{\rm px}$ can be considered as upper limits. We consider these conservatives values as our final uncertainty on the filament orientation, displayed in Table~\ref{tab:uncertainties}.

We can see that the precision is very good, $\leq 0.2^{\circ}$ for ideal filaments and $\leq 1.8^{\circ}$ for Plummer-type filaments for the set of simulations SimSet~1 (default noise level and short aspect ratio). These uncertainties increase with increasing noise (SimSet~2), up to $\leq 0.4^{\circ}$ (ideal) and $\leq 3.1^{\circ}$ (Plummer). The impact of noise level on the filament orientation estimate is much less important than the one observed for the width, with a degradation of only a factor of $\leq 2$ for both kind of synthetic filament profiles.

The uncertainties decrease with increasing aspect ratio (SimSet~3), down to $\leq 0.1^{\circ}$ (ideal) and $\leq 1.2^{\circ}$ (Plummer), which is again in accordance with our expectation that more elongated filaments are easier to recognize.

\begin{table*}\centering 
\caption{Uncertainties of {\tt FilDReaMS} obtained in three different cases ({\bf left}) in the estimation of width ({\bf center}) and orientation ({\bf right}).}

\begin{tabular}{m{3cm} | m{0.01cm} m{2cm} m{2cm} m{0.01cm} | m{0.01cm} m{2cm} m{2cm} m{0.01cm}}
\toprule\toprule
\textbf{\large SimSet ID} & & \multicolumn{2}{c}{Width} & & & \multicolumn{2}{c}{Orientation} & \\ 
& & \centering Ideal & \centering Plummer & & & \centering Ideal & \centering Plummer & \\ 

\midrule

SimSet~1 & & \centering $\simeq 0.01\,{\rm px}$ & \centering $\simeq 0.4\,{\rm px}$ & & & \centering $\leq 0.2^{\circ}$ & \centering $\leq 1.8^{\circ}$ & \\

\midrule

SimSet~2 & & \centering $\simeq 0.4\,{\rm px}$ & \centering $\simeq 2.7\,{\rm px}$ & & & \centering $\leq 0.4^{\circ}$ & \centering $\leq 3.1^{\circ}$ & \\

\midrule

SimSet~3 & & \centering $\simeq 0.002\,{\rm px}$ & \centering $\simeq 0.2\,{\rm px}$ & & & \centering $\leq 0.1^{\circ}$ & \centering $\leq 1.2^{\circ}$ & \\

\bottomrule\bottomrule
\end{tabular}
\label{tab:uncertainties}
\end{table*}

\section{First astrophysical application and comparison with previous work}
\label{sec:comparison}

\begin{figure*}
    \centering
    \includegraphics[width=\textwidth]{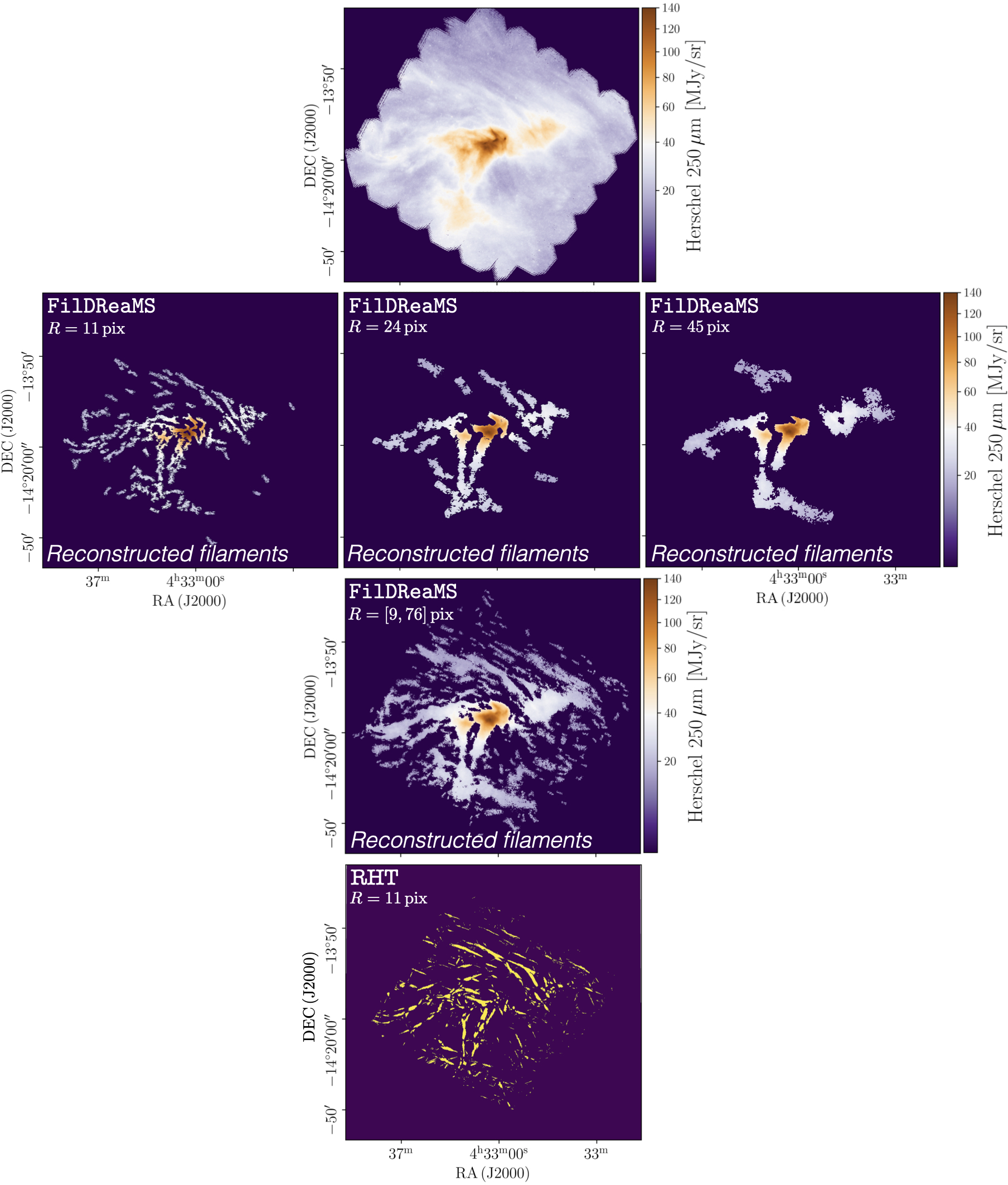}
    \caption{Application of {\tt FilDReaMS} to the $250\,\mu m$ intensity map of the {\it Herschel} G210 field.
    {\bf Top row:} $250\,\mu m$ intensity map of G210.
    {\bf Second row:} Reconstructed filaments with bar widths $W_{\rm b} = 6\,{\rm px}$ ({\bf left}), $W_{\rm b} = 10\,{\rm px}$ ({\bf middle}), and $W_{\rm b} = 17\,{\rm px}$ ({\bf right}).
    {\bf Third row:} Map of all the reconstructed filaments with bar width in the range $[5, 27]\,{\rm px}$.
    {\bf Bottom row:} Filaments detected with {\tt RHT} using a 2D top-hat radius $R=11\,{\rm px}$ \citep[][]{Malinen_2016}.
    The color code of the reconstructed filaments indicates their column density $N_{\rm H_2}$.}
    \label{fig:G210_previous_results}
\end{figure*}

As a first astrophysical illustration, we apply {\tt FilDReaMS} to the {\it Herschel} $250\,{\rm \mu m}$ intensity map of the G210 field (top panel of Fig.~\ref{fig:G210_previous_results}). The left, middle, and right panels in the second row of Fig.~\ref{fig:G210_previous_results} show the networks of reconstructed filaments with bar widths $W_{\rm b} = 6\,{\rm px}$, $12\,{\rm px}$, and $17\,{\rm px}$. The corresponding radius of the 2D top-hat kernel used to filter out the large scales in the initial map (to obtain map B; see top middle panel of Fig.~\ref{fig:FilDReaMS_method}) is $R=11\,{\rm px}$, $24\,{\rm px}$, and $45\,{\rm px}$, respectively. We immediately see that the majority of the small filaments are sub-structures of larger ones, while only a small fraction of them covers areas not already included in larger structures, stressing the hierarchical organisation of these filamentary structures in the interstellar medium.

The map of all the reconstructed filaments with bar widths between $W_{\rm b} = 5\,{\rm px}$ and $27 \,{\rm px}$ is displayed in the third row of Fig.~\ref{fig:G210_previous_results}.
This map bears some obvious resemblance to the initial map (top row), while a direct by-eye inspection indicates that {\tt FilDReaMS} recovers most of the filamentary structures independently of the intensity.

The striations in the northern-central part of the region are better reconstructed (i.e., over longer path lengths) when integrating all bar widths than for any single value of $W_{\rm b}$. These striations are in fact composed of filaments of different bar widths, mostly parallel to each other, with the thinner corresponding to the crests of the larger.

We now compare the above results to those of \citet[][]{Malinen_2016}, who applied the {\tt RHT} method to the {\it Herschel} $250\,{\rm \mu m}$ intensity map of G210 (top panel of Fig.~\ref{fig:G210_previous_results}), using a spatial filtering with $R=11\,{\rm px}$.
The filaments detected with {\tt RHT} are displayed in the bottom row of Fig.~\ref{fig:G210_previous_results}, where they can be compared to the filaments detected with {\tt FilDReaMS} (left panel in the second row).
It appears that both methods yield similar filamentary networks, with however a few noticeable differences: {\tt RHT} detects more filaments in the most diffuse part of the region, while in the densest part {\tt FilDReaMS} detects more strands and structures branching out from the main central filament.
The long and thin striations that seem to escape detection with {\tt FilDReaMS} for $R=11\,{\rm px}$ could be missed because of the short aspect ratio of the {\tt FilDReaMS} model bar. However, they are recovered when considering a short range of values around $R=11\,{\rm px}$.

\section{Concluding remarks}
\label{sec:conclusion}

In this paper, we presented a new method to detect filaments of different widths in a map of intensity (in a broad sense) $\I$.
We called our method {\tt FilDReaMS}, for {\bf Fil}ament {\bf D}etection and {\bf Re}construction {\bf a}t {\bf M}ultiple {\bf S}cales. In brief, {\tt FilDReaMS} uses a rectangular model bar of width $W_{\rm b}$ and aspect ratio $r_{\rm b}$,
where $W_{\rm b}$ is meant to cover a broad range of values and $r_{\rm b}$ is a free parameter (typically set to $r_{\rm b} = 3$).
For any given value of $W_{\rm b}$, {\tt FilDReaMS} (1) detects potential filaments that can be locally approximated by the model bar, (2) retains the potential filaments with significance $S>1$ (Eq.~\ref{eq:criterion_significance}), (3) reconstructs the true shape and the intensity of physical filaments from the initial map of $\I$ together with the associated binary map (see Fig.~\ref{fig:filament_reconstruction}), and (4) assigns a filament orientation angle, $(\psi_{\rm f}^{\star})_i$, to each pixel $i$ of a reconstructed filament of width $W_{\rm b}$. 
After repeating the procedure for all the values of $W_{\rm b}$, a most significant bar width, $(W_{\rm b}^\star)_i$, is derived for each pixel $i$ of all the reconstructed filaments, and the most prevalent bar widths of the map, $W_{\rm b}^{\star{\rm peak}}$, are inferred from the peaks of the histogram of $W_{\rm b}^\star$.
The $W_{\rm b}^{\star{\rm peak}}$ can then be cautiously converted to the widths of the often-used Plummer-type profiles, $2R_{\rm flat}$ (see Fig.~\ref{fig:scale_plum_rect} and Table~\ref{tab:linear_fits}).

Thus {\tt FilDReaMS} makes it possible to detect filaments of a given bar width, to identify the most prevalent bar widths (and corresponding Plummer widths) in a given map, and to derive the local orientation angles of the detected filaments. The main assets of {\tt FilDReaMS} are
\begin{itemize}
    \item the ability to detect filaments over a broad range of widths;
    \item the small number of free parameters (only three; see Table~\ref{tab:FilDReaMS_parameters});
    \item the speed of execution: typically, for a given map, running {\tt FilDReaMS} takes about $20-30\,{\rm sec}$ for each value of $W_{\rm b}$ and roughly $10-20\,{\rm min}$ to cover the entire range of $W_{\rm b}$;
    \item the user-friendliness, which makes it particularly suited for statistical studies. 
\end{itemize}

{\tt FilDReaMS} opens wide perspectives of application to astrophysical data, in order to study the processes of the stellar formation inside the interstellar filamentary structures. To start with we investigate the interplay between the orientation of the filaments and the Galactic magnetic field in a companion paper \citep{Carriere_2022b}, generalising the analysis of \citet{Malinen_2016} to four other fields.  
A broader statistical analysis over 116 {\it Herschel} fields is also in preparation. 

\begin{acknowledgements}
We extend our deepest thanks to our referee, Gina Panopoulou, for her careful reading of our paper and for her many constructive comments and suggestions.
We also acknowledge useful discussions with Dana Alina, Susan Clark, Mika Juvela, and Julien Montillaud. 
Herschel SPIRE has been developed by a consortium of institutes led by Cardiff University (UK) and including University Lethbridge (Canada); NAOC (China); CEA, LAM (France); IFSI, University Padua (Italy); IAC (Spain); Stockholm Observatory (Sweden); Imperial College London, RAL, UCL-MSSL, UKATC, University Sussex (UK); Caltech, JPL, NHSC, University Colorado (USA). This development has been supported by national funding agencies: CSA (Canada); NAOC (China); CEA, CNES, CNRS (France); ASI (Italy); MCINN (Spain); SNSB (Sweden); STFC (UK); and NASA (USA).
\end{acknowledgements}

\bibliographystyle{aa}
\bibliography{biblio}

\end{document}